\documentclass[journal=jpcafh,manuscript=article]{achemso}
\usepackage{longtable}
\usepackage{multirow}
\usepackage{amsmath}
\usepackage{subfigure}
\usepackage[usenames,dvipsnames]{color}
\usepackage{soul}
\usepackage{bm}

\usepackage{xr}
\usepackage{ifpdf}
\usepackage{amssymb}
\usepackage{titlecaps}
\Addlcwords{a an the of on at with for and in into from by without}

\makeatletter
\newcommand*{\addFileDependency}[1]{% argument=file name and extension
  \typeout{(#1)}
  \@addtofilelist{#1}
  \IfFileExists{#1}{}{\typeout{No file #1.}}
}
\makeatother

\newcommand*{\myexternaldocument}[1]{%
    \externaldocument{#1}%
    \addFileDependency{#1.tex}%
    \addFileDependency{#1.aux}%
}
\myexternaldocument{si}

\usepackage{multicol} \usepackage{wrapfig} \usepackage{enumitem}
\usepackage{bm} \usepackage{outlines} %for itemize with subitems
\SectionNumbersOn

\author{Seyedeh Maryam Salehi, Debasish Koner}
\affiliation[University of Basel]{Department of Chemistry, University
  of Basel, Klingelbergstrasse 80 , CH-4056 Basel, Switzerland.}
\author{Markus Meuwly}
\affiliation[University of Basel]{Department of Chemistry, University
  of Basel, Klingelbergstrasse 80 , CH-4056 Basel, Switzerland.}
\email{m.meuwly@unibas.ch}

\title{The Dynamics and Infrared Spectrocopy of Monomeric and Dimeric
  Wild Type and Mutant Insulin}

\begin{document}

\date{\today}

\begin{abstract}
The infrared spectroscopy and dynamics of -CO labels in wild type and
mutant insulin monomer and dimer are characterized from molecular
dynamics simulations using validated force fields. It is found that
the spectroscopy of monomeric and dimeric forms in the region of the
amide-I vibration differs for residues B24-B26 and D24-D26, which are
involved in dimerization of the hormone. Also, the spectroscopic
signatures change for mutations at position B24 from phenylalanine -
which is conserved in many organisms and known to play a central role
in insulin aggregation - to alanine or glycine. Using three different
methods to determine the frequency trajectories - solving the nuclear
Schr\"odinger equation on an effective 1-dimensional potential energy
curve, instantaneous normal modes, and using parametrized frequency
maps - lead to the same overall conclusions. The spectroscopic
response of monomeric WT and mutant insulin differs from that of their
respective dimers and the spectroscopy of the two monomers in the
dimer is also not identical. For the WT and F24A and F24G monomers
spectroscopic shifts are found to be $\sim 20$ cm$^{-1}$ for residues
(B24 to B26) located at the dimerization interface. Although the
crystal structure of the dimer is that of a symmetric homodimer,
dynamically the two monomers are not equivalent on the nanosecond time
scale. Together with earlier work on the thermodynamic stability of
the WT and the same mutants it is concluded that combining
computational and experimental infrared spectroscopy provides a
potentially powerful way to characterize the aggregation state and
dimerization energy of modified insulins.
\end{abstract}

\section{Introduction}
Insulin is a small, aggregating protein with an essential role in
regulating glucose uptake in cells. Physiologically, it binds to the
insulin receptor (IR) in its monomeric form but thermodynamically the
dimer is more stable for the wild type (WT)
protein.\cite{Strazza1985,TidorJMB1994,ZoeteProt2005} The storage form
is that of a zinc-bound hexamer with either two or four Zn
atoms.\cite{BakerPhilos1988} Hence, to arrive at the functionally
relevant monomeric stage, insulin has to cycle through at least two
dissociation steps: from the hexamer to three dimers and from the
dimer to the monomer.\\

\noindent
For pharmacological applications the dimer$\leftrightarrow$monomer
equilibrium is particularly relevant because for safe insulin
administration this equilibrium needs to be tightly
controlled. However, reliable experimental physico-chemical
information about the relative stabilization of insulin monomer and
dimer, which is $-7.2$ kcal/mol in favour of the
dimer,\cite{Strazza1985} is only available for the WT and the barrier
between the two states is unknown. For mutant insulins, there is no
such quantitative information from experiments. On the other hand,
insulin has become a paradigm for studying coupled folding and
binding,\cite{tokmakoff:2020} whether or not association proceeds
along one or multiple pathways,\cite{bagchi:2018,dinner:2020} and for
the role of water in protein
association.\cite{MM.insulin:2018,he:2019,bagchi:2020} Most of these
studies were based on atomistic molecular dynamics (MD) simulations
and provided remarkable insight into functionally relevant processes
for this important system.\\

\noindent
Infrared spectroscopy has been proposed\cite{Suydam.halr.sci.2006} and
recently demonstrated\cite{vibvtark:PCCP2017} to provide a way to
quantify protein-ligand binding strengths through observation of
spectroscopic shifts. The physical foundation for this is the Stark
effect which is based on the electrostatic interaction between a local
reporter and the electric field generated by its environment. Using
accurate multipolar force fields\cite{bereau2013} it was possible to
assign the structural substates in photodissociated CO from
Myoglobin\cite{BJ2008} whereas more standard, point charge-based force
fields are not suitable for such
investigations.\cite{nutt-biophys-03}\\

\noindent
The frequency trajectory of a local reporter can be followed in
different ways. One of them uses so-called parametrized ``frequency
maps'' which are precomputed for a given reporter from a large number
of ab initio
calculations.\cite{skinner-map-JPCB2011,knoester-jpc-model-2006,knoester-jpc-erratum-2012,Tokmakoff-map-jcp-2013}
Alternatively, the sampling of the configurations and computing
frequencies for given snapshots can also be done using the same energy
function (``scan''). In this approach, the MD simulations are carried
out with the same energy function that is also used for the analysis,
which is typically a multipolar representation for the electrostatics
around the spectroscopic probe and an anharmonic (Morse) for the
bonded terms.\cite{Lee13p054506,NMA2dir-MM-2014} On each snapshot, the
local frequency is determined from either an instantaneous normal mode
(INM) calculation or by solving the 1D or three-dimensional nuclear
Schr\"odinger equation.\cite{Salehi:N3-JPCB2019}\\

\noindent
Here, the WT proteins and two mutants at position B24 (Phe) are
considered. Phenylalanine B24 is located at the dimerization interface
and invariant among insulin sequences.\cite{HuaNature1991} Compared
with the WT, the SerB24,\cite{TagerPNAS1983,HanedaPNAS1985}
LeuB24,\cite{TagerPNAS1980} and HisB24\cite{Brzozowski2013} analogues
show reduced binding potency towards the receptor. On the other hand,
substitutions such as GlyB24, D-AlaB24, or D-HisB24 are well tolerated
as judged from their binding affinity. Nevertheless, substitutions
such as GlyB24 (F24G) or AlaB24 (F24A) were found to have reduced
stability of the modified insulin dimer, both from simulations and
experiment,\cite{MM.insulin:2018,Chen2000,DeFelippis2001} and these
are the variants considered in the present work.\\

\noindent
In the present work the infrared spectrum in the amide-I stretch
region is studied for wild type (WT) and two mutant insulins in their
monomeric and dimeric states using accurate multipolar force
fields. The IR lineshapes are calculated from frequency trajectories
calculated by using a normal mode analysis, solving the Schr\"odinger
equation from a 1-d scan along the amide-I normal mode and using
previously parametrized maps. First, the methods are presented. Then,
results for IR lineshapes and frequency correlation functions from
scanning along the amide-I normal mode are presented and discussed and
compared with the two other approaches. Finally, conclusions are
drawn.\\

\section{Methods}

\subsection{Molecular Dynamics Simulations}
All molecular dynamics (MD) simulations were carried out using the
CHARMM\cite{Charmm-Brooks-2009} package together with
CHARMM36\cite{MacKerell1998} force field including the CMAP
correction\cite{brooks:2004,brooks:2004.2} and multipoles up to
quadrupole on the [CONH]-part of the
backbone.\cite{NMA2dir-MM-2014,NMA-MM-2015} The X-ray crystal
structure of the insulin dimer was solvated in a cubic box ($75^3$
\AA\/$^3$) of TIP3P\cite{TIP3P-Jorgensen-1983} water molecules, which
leads to a total system size of 40054 atoms. For the monomer
simulations, chains A and B were retained and also solvated in a water
box ($75^3$ \AA\/$^3$ ), the same box size as the dimer. In these
simulations the
multipolar\cite{Kramer2012,bereau2013,NMA2dir-MM-2014,NMA-MM-2015}
force field is used for the entire amide groups and all CO bonds are
treated with a Morse potential $V(r) =D_e
(1-\exp(-\beta(r-r_e)))^2$. The parameters are $D_e = 141.666$
kcal/mol, $\beta = 2.112$ \AA$^{-1}$ and $r_0 = 1.231$ \AA\/.\\

\begin{figure}[H]
\begin{center}
\includegraphics[width=0.8\textwidth]{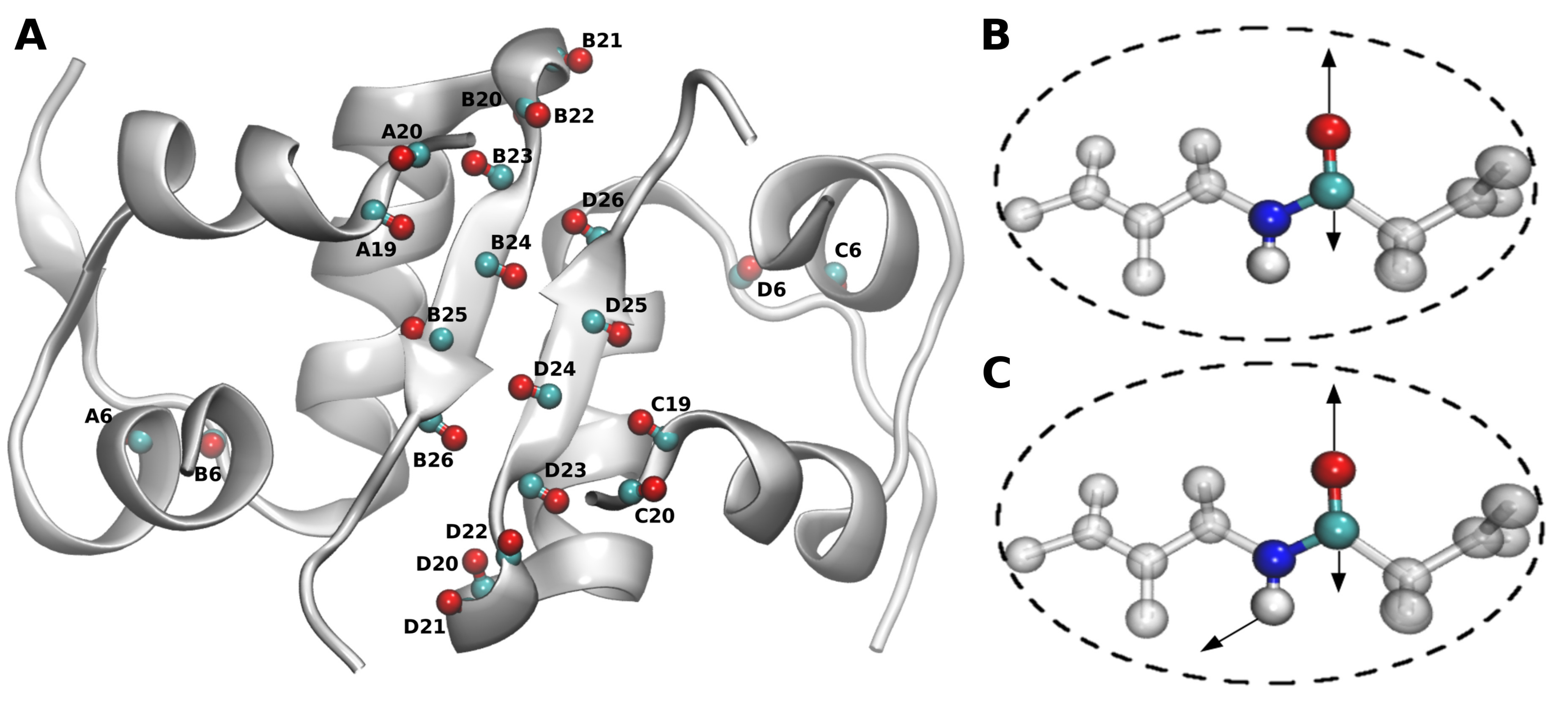}
\caption{Panel A: Structure of wild type insulin dimer with the -CO
  labels that are specifically probed in the present work. The
  dimerization interface involves residues B24-B26 and D24-D26. Panels
  B and C show the displacement vectors for the two scan approaches
  considered to construct 1D potentials along the CO and CONH
  directions, respectively.}
\label{fig:structure}
\end{center}
\end{figure}

\noindent
Hydrogen atoms were included and the structures of all systems were
minimized using 2000 steps of steepest descent (SD) and 200 steps of
Newton Raphson (ABNR) followed by 20 ps of equilibration MD at 300
K. A Velocity Verlet integrator\cite{Swope:1982} and Nos\'e-Hoover
thermostat\cite{Nose:1984,Hoover:1985} were employed in the $NVT$
simulations. Then production runs (1 ns or 5 ns) were carried out in
the $NpT$ ensemble, with coordinates saved every 10 fs for subsequent
analysis. For the $NpT$ simulations an Andersen and Nos\'e-Hoover
constant pressure and temperature algorithm was
used\cite{Andersen:1980,Nose:1983,Hoover:1985} together with a
leapfrog integrator.\cite{Hairer2003} a coupling strength for the
thermostat of 5 ps and a damping coefficient of 5 ps$^{-1}$. All bonds
involving hydrogen atoms were constrained using
SHAKE\cite{SHAKE-Gunsteren-1997}.  Nonbonded interactions were treated
with a switching function\cite{Steinbach1994} between 10 and 14 \AA\/
and for the electrostatic interactions, the Particle Mesh Ewald (PME)
method was used with grid size spacing of 1 \AA\/, characteristic
reciprocal length $\kappa = 0.32$ \AA\/$^{-1}$, and interpolation
order 4.\cite{Darden1993} Figure \ref{fig:structure}A shows the
insulin dimer highlighting some of the CO labels studied in the
current work with particular attention to the -CO labels at the
protein-protein interface (B24-B26) and (D24-D26).\\

\subsection{Frequencies from Solving the 1d Schr\"odinger Equation: Scan}
Anharmonic transition frequencies can be determined from calculating
the 1-d potential energy along the CO or amide-I normal mode (from a
normal mode analysis on N-methyl acetamide (NMA) in the gas phase) and
solving the nuclear Schr\"odinger equation (SE) for each snapshot
using a discrete variable representation (DVR)
approach\cite{col92:1982}. It was shown previously for
NMA\cite{MM.rev.jcp:2020} that frequency trajectories obtained from
solving the SE on the 1-d PES scanned along either the CONH (amide-I)
or the CO mode (see Figure \ref{fig:structure}B and C) result in
similar decay times with frequencies shifted by some $\sim 15$
cm$^{-1}$. Here, scans were performed for each snapshot for 61 points
along the CO normal mode vector around the minimum energy structure
using the same energy function as that used for the MD simulations,
i.e.  a multipolar representation of the electrostatics and an
anharmonic Morse potential for the CO-bond. An RKHS representation of
the 1-d PES is then constructed from these energies and the SE is
solved on a grid ($-0.53$ \AA\ $ <r< 0.53$ \AA\ ) using a reduced mass
of 1 amu.\cite{MM.rev.jcp:2020} For direct comparison, scans along the
amide-I mode were also carried out for selected residues.\\

\subsection{Instantaneous Normal Mode}
The instantaneous (harmonic) frequencies for each snapshot of the
trajectory from the $NPT$ simulation were calculated for the same
snapshots for which the scan along the CO normal mode was carried out,
see above. Such instantaneous normal modes (INM) are determined by
minimizing CO or [CONH] while keeping the environment (protein plus
solvent) fixed. Next, normal modes were calculated from the ``vibran''
facility in CHARMM.\\

\subsection{The Amide I Frequency Maps}
The frequency map used in the present work is that parametrized by
Tokmakoff and coworkers.\cite{Tokmakoff-map-jcp-2013} It requires MD
simulations to be run with fixed CO bond length and is based on the
expression
\begin{equation}
\label{eq:3}
    \omega_i= \omega_0 + aE_{C_{i}}+E_{{N}_i}
\end{equation}
where $\omega_i$ is the instantaneous frequency for the $i$th
vibrational label, $E_{C_{i}}$ is the electric field on the C atom in
the $i$th label along the C=O bond direction, and $E_{N_{i}}$ is that
on the N atom. Parameters $\omega_0$, $a$, and $b$ were fitted such
that they optimally reproduce the experimental IR absorption spectra
of NMAD. The optimized backbone map is\cite{Tokmakoff-map-jcp-2013}
\begin{equation}
\label{eq:tok}
    \omega_i=1677.9+2557.8E_{\rm C_{i}}-1099.5E_{N_{i}}
\end{equation}
In this equation, $\omega_i$ is in cm$^{-1}$ and $E_{C_{i}}$ and
$E_{N_{i}}$ are in atomic units.\\

\subsection{Frequency Fluctuation Correlation Function and Lineshape}
From the harmonic or anharmonic frequency trajectory $\omega_i(t)$ or
$\nu_i(t)$ for label $i$ its frequency fluctuation correlation
function, $\langle \delta \omega(0) \delta \omega(t) \rangle$ is
computed. Here, $\delta \omega(t) = \omega(t) - \langle \omega(t)
\rangle$ and $\langle \omega(t) \rangle$ is the ensemble average of
the transition frequency. From the FFCF the line shape function 
\begin{equation}\label{eq:5}
g(t) = \int_{0}^{t} \int_{0}^{\tau^{'}} \langle \delta
\omega(\tau^{''}) \delta \omega(0) \rangle d\tau^{''} d\tau^{'}.
\end{equation}
is
determined within the cumulant approximation. To compute $g(t)$, the FFCF is numerically
integrated using the trapezoidal rule and the 1D-IR spectrum is
calculated according to\cite{2DIRbook-Hamm-2011}
\begin{equation}
I(\omega) = 2 \Re \int^\infty_0
e^{i(\omega-\langle\omega\rangle)t} e^{-g(t)} e^{-\frac{t \alpha}{2T_1}}dt
\end{equation}

where $\langle\omega\rangle$ is the average transition frequency
obtained from the distribution, $T_1 =0.45$ ps is the vibrational
relaxation time and $\alpha = 0.5$ is a phenomenological factor to account for lifetime broadening.\cite{2DIRbook-Hamm-2011}\\

\noindent
For extracting time information from the FFCF, $\langle \delta
\omega(t) \delta \omega(0) \rangle$ is fitted to an empirical
expression\cite{hynes:2004}
\begin{equation}
  \langle \delta \omega(t) \delta \omega(0) \rangle = a_{1}
  \cos(\gamma t) e^{-t/\tau_{1}} + \sum_{i=2}^{n} a_{i}
  e^{-t/\tau_{i}} + \Delta_0
\label{eq:ffcf}
\end{equation}
where $a_i$ are amplitudes, $\tau_i$ are decay times and $\Delta_0$ is
an offset for long correlation times. The $\cos-$term allows to
capture a short-time recurrence (anticorrelation) that may or may not
be present in the correlation function. This minimum at very short
time ($t \sim 0.1$ ps) is known from previous
simulations\cite{skinner:2006} and can be related to the strength of
the interaction between solute and
solvent\cite{hynes:2004,Lee13p054506,NMA2dir-MM-2014,Salehi:N3-JPCB2019}
or between the spectroscopic probe and its environment (as in the
present case). The decay times $\tau_i$ of the frequency fluctuation
correlation function reflect the characteristic time-scale of the
solvent fluctuations to which the solute degrees of freedom are
coupled. In most cases the FFCFs were fitted to an expression
containing two decay times using an automated curve fitting tool from
the SciPy library.\cite{2020SciPy-NMeth} Only if the quality of the
resulting fit was evidently insufficient, a third decay time was
included.\\

\section{Results}
The results section is structured as follows. First, a brief account
is given of representative structures along the trajectories for the
different simulation conditions used. Next, the amide-I spectroscopy
for the WT monomer and dimer using the ``scan'' approach is
given. This is followed by the spectroscopy for the mutant monomer and
dimer compared with the WT systems. Then, a comparative discussion of
the results for WT and mutant monomer and dimer is given for the three
methods to determine the frequency trajectories (``scan'', ``INM'' and
``map'') and finally, the FFCFs from the ``scan'' and ``INM''
frequency trajectories are discussed.\\

\subsection{Structural Characterization}
\begin{figure}[H]
\begin{center}
\includegraphics[width=0.60\textwidth,angle=-90]{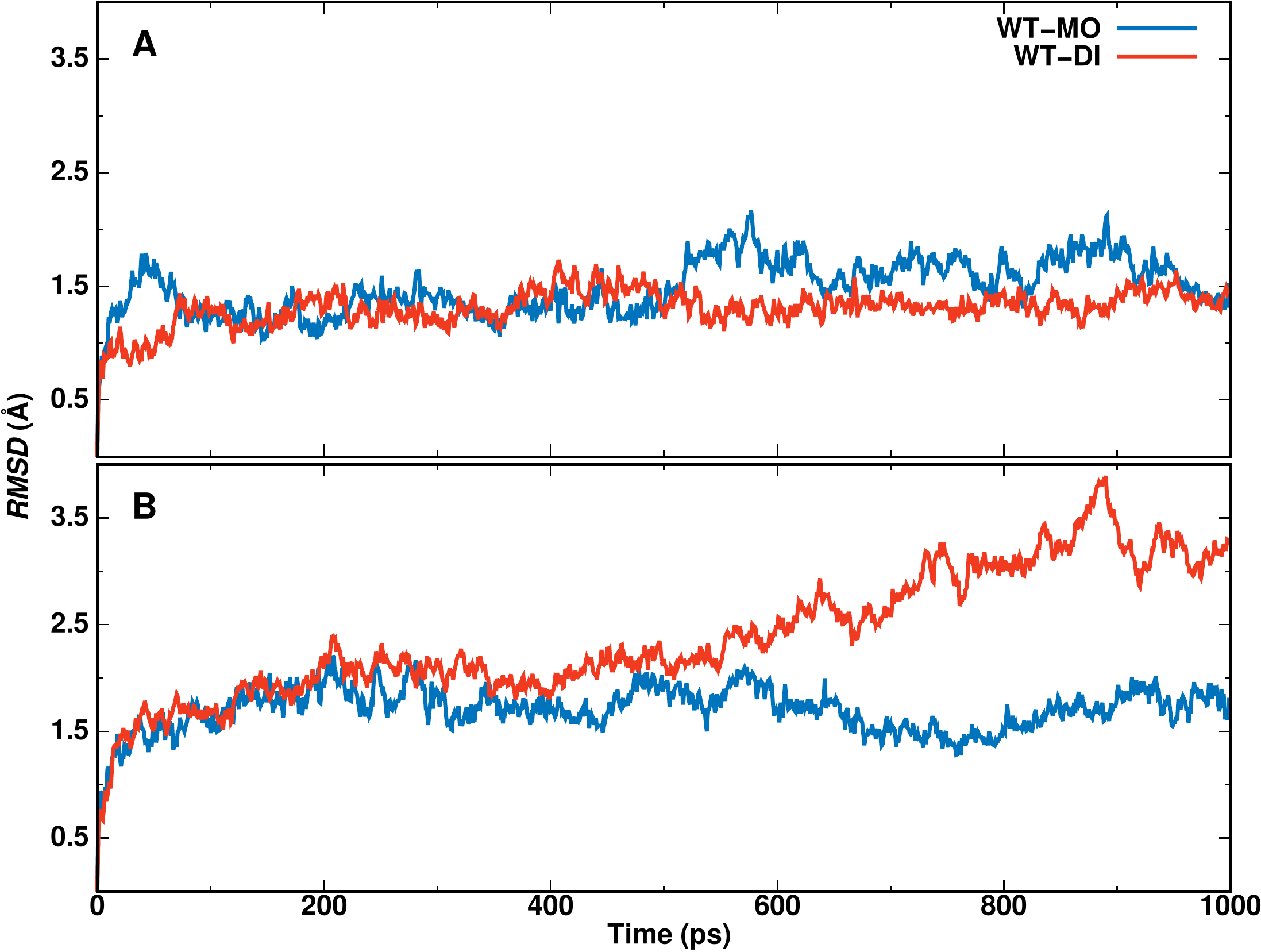}
\caption{The structural RMSD between the reference X-ray structure and
  the Wild type monomer and dimer insulin for A) flexible and B) constrained
  CO.}
\label{fig:rmsd}
\end{center}
\end{figure}

The root mean squared deviation between the reference X-ray structure
and those of the monomer and dimer structure of the WT protein in
solution is reported in Figure \ref{fig:rmsd}. Typically, the RMSD is
around 1.5 \AA\/ which is indicative of a stable simulation on the
nanosecond time scale. Such RMSD values have also been reported from
simulations in smaller water
boxes.\cite{MeuwlyJMB2004,ZoeteProt2005}\\
  
\noindent
With constrained CO (as is required for using the frequency maps) the
structure of the monomer is equally well maintained whereas for the
dimer it starts to deviate from the reference structure by $\sim 3$
\AA\/ after 0.8 ns. This is indicative of structural changes which
involve separation of the terminal of chain B (PheB1 and AlaB30) from
each other. A similar but less pronounced effect was also observed for
chain D between PheD1 and AlaD30.\\

\subsection{Amide-I Spectroscopy Using Scan for WT and Mutant Monomer and Dimer}
\begin{figure}[H]
\begin{center}
\includegraphics[width=0.7\textwidth,angle=-90]{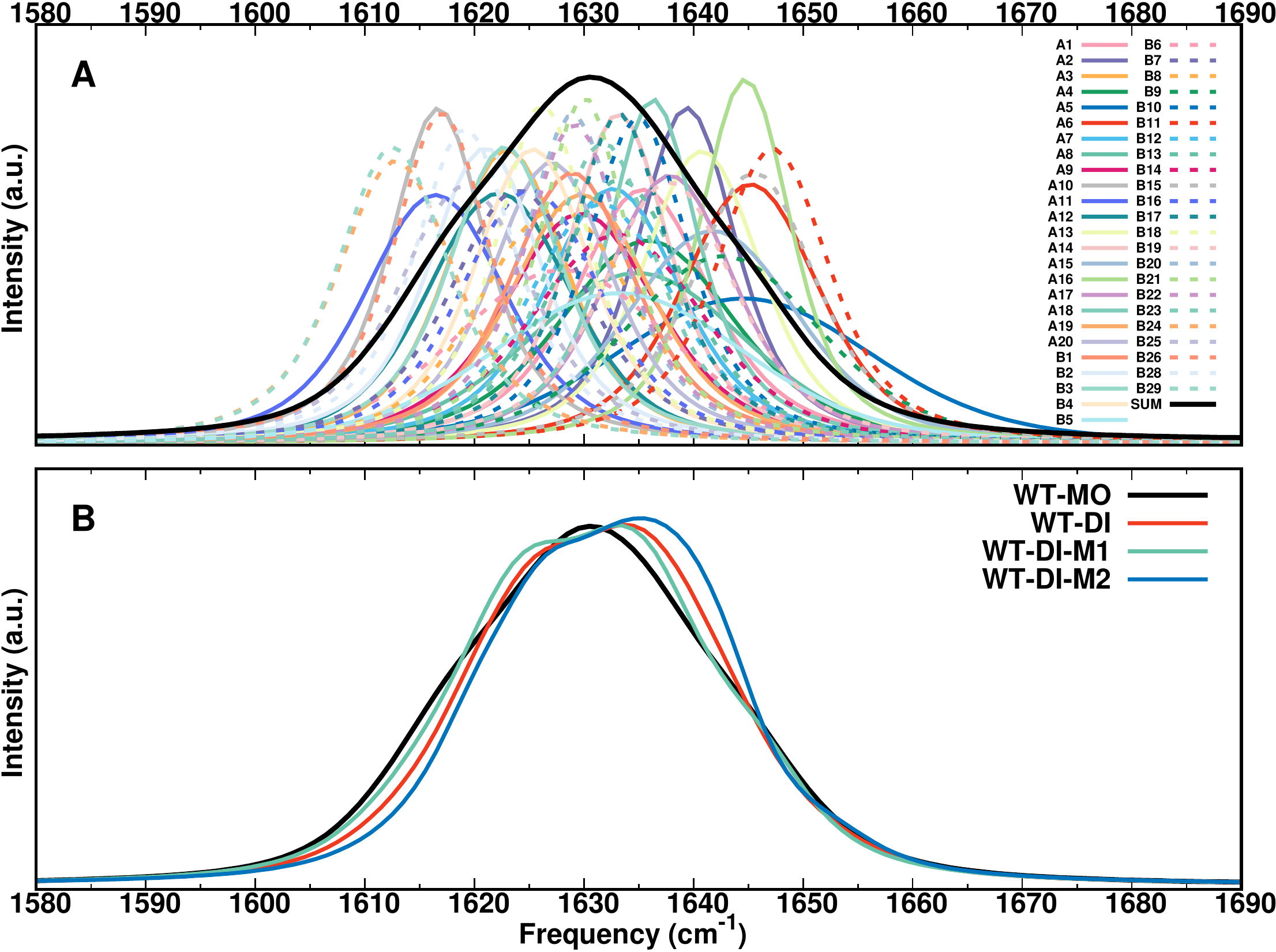}
\caption{Panel A: 1D IR spectra for all residues in WT monomer based on
  ``scan'' for the frequency calculation. The labels for the
  individual line shapes are given in the panel and the overall sum is
  the solid black line. Panel B: The total lineshape for all CO probes
  of the monomer (black) compared with that of M1 (green) and M2
  (blue) within the dimer and with the dimer itself (red). All
  lineshapes are scaled to the same maximum intensity. The line shapes
  are determined from 1 ns simulations and the snapshots analyzed are
  separated by 10 fs.}
\label{fig:scan.monomer}
\end{center}
\end{figure}

To set the stage, the Amide-I spectroscopy for the WT monomer and
dimer is discussed from frequency trajectories obtained by scanning
along the CO normal mode for each snapshot. Figure
\ref{fig:scan.monomer} reports the lineshapes for all CO-labels for
the WT monomer. Lineshapes for chain A are solid lines and those for
chain B are dashed. The overall lineshape for the monomer (black solid
line) is centered at 1630.5 cm$^{-1}$ and has a full width at half
maximum of $\sim 30$ cm$^{-1}$, compared with a center frequency of
$\sim 1650$ cm$^{-1}$ and a FWHM of $\sim 30$ cm$^{-1}$ from
experiments.\cite{tokmakoff:2016,tokmakoff:2016.2} When comparing the
position of the frequency maximum it should be noted that the present
parametrization is for NMA and slight readjustments of the Morse
parameters could be made to yield quantitative agreement. However, for
the present purpose such a step was deemed unnecessary.\\

\noindent
On the other hand, scanning the 1-dimensional potential along the
amide-I normal mode shifts the frequencies by about 30 cm$^{-1}$ to
the blue (see Figure S1A). The correlation
between scanning along the CO and amide-I normal modes is high, as
Figure S1C shows. In addition, the full 1D
infrared spectrum was also calculated from scanning along the amide-I
normal mode (Figure S2) and confirms the overall
shift to the blue by 25 cm$^{-1}$ while maintaining the shape and
width of the total lineshape from scanning along the CO normal mode.\\

\noindent
Most notably, the center frequencies for each of the labels cover a
range from 1612.5 cm$^{-1}$ (residues B24, B29) to 1647.5 cm$^{-1}$
(residue B11) although the bonded potential (Morse) for the CO stretch
is the same for all 51 labels. Hence, the multipolar charge
distribution used for the electrostatics and its interaction with the
environment leads to the displacements of the center frequencies. The
linewidths also vary for the -CO probes at the different locations
along the polypeptide chain and cover a range from 10 cm$^{-1}$
(Residues A10, A16, A18, B18, B21) to 28 cm $^{-1}$ (Residue A5).\\

\begin{figure}[H]
\begin{center}
\includegraphics[width=0.99\textwidth]{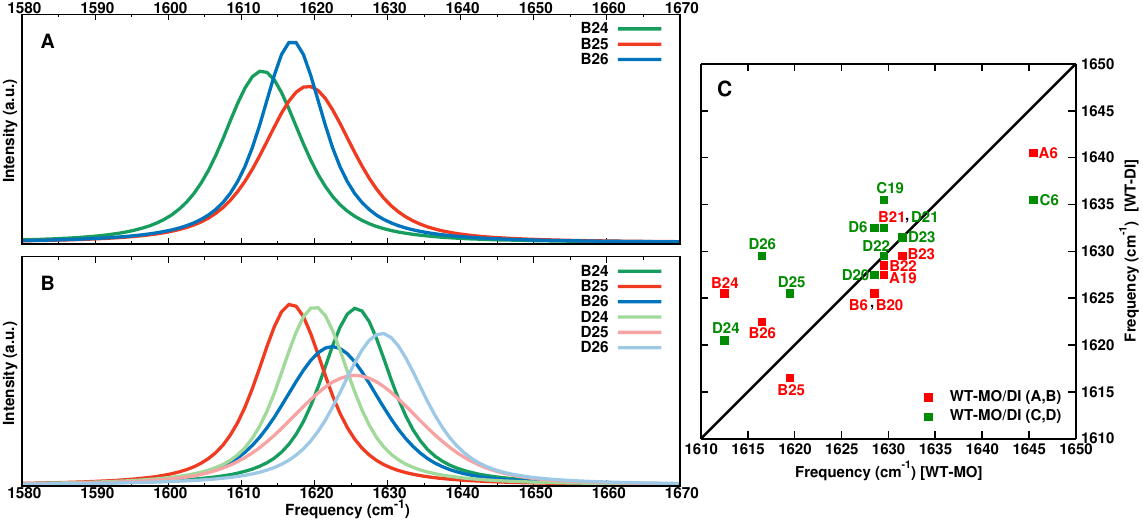}
\caption{1D IR spectra for WT monomer (panel A) and dimer (panel B)
  for residues at the dimerization interface (B24-B26) and (B24-B26,
  D24-D26), respectively, based on ``scan'' for frequency
  calculation. Panel C compares the maximum frequency of the 1D IR
  spectra for the selected residues (A6, A19, B6, B20-B26, C6, C19,
  D6, D20-D26) between WT monomer and dimer.}
\label{fig:scan.modi}
\end{center}
\end{figure}

\noindent
Selected lineshapes for the monomer and each of the two monomers
within insulin dimer from scanning along the CO normal mode are
reported in Figures \ref{fig:scan.modi} and S3
and the individual and total lineshapes for the two monomers (M1 and
M2) within the dimer are shown in Figures S4 and
S5. For the dimer it is noted that some probes
at symmetry related positions within the dimer structure typically
have their maxima at different frequencies. In other words,
structurally related -CO probes sample different environments in the
hydrated system at room temperature. The overall lineshapes of M1 and
M2 are directly compared with that of the isolated monomer and the
dimer in Figure \ref{fig:scan.monomer}B. The lineshape of M1 and M2
differ which confirms the asymmetry noted earlier from X-ray
experiments.\cite{BakerPhilos1988,falconi:2001} Also, the spectroscopy
of the isolated monomer differs from that of M1 and M2 within the
dimer. Notably, the -CO groups involved in the hydrogen bonding motif
of the insulin dimer (B24 to B26 and D26 to D24) display frequency
maxima that differ by $\sim 10$ cm$^{-1}$. Other -CO reporters, such
as B20 and D20, have their maxima only $\sim 5$ cm$^{-1}$ apart.\\

\noindent
It is also observed that the absolute frequency maximum of the same
reporter in the monomer and in the dimer can differ. For example,
while the maximum frequency of -CO at position B24 in the monomer is
at 1612.5 cm$^{-1}$ the maxima for B24 and D24 in the dimer are at
1625.5 cm$^{-1}$ and 1620.5 cm$^{-1}$. Hence, in addition to a
splitting in the dimer spectrum also an overall shift of the
frequencies compared with the monomer is found. Again, these effects
are largest for the dimerization motif and for residues A/C6, see
Figure \ref{fig:scan.modi}C.\\

\noindent
The close agreement of the computed overall spectrum with the
experimentally measured one (see above) and the fact that the same
computational model was successful in describing the spectroscopy and
dynamics of hydrated NMA\cite{NMA2dir-MM-2014,desmond-jcpb-2019}
provides a meaningful validation of the present approach.\\

\noindent
{\it Amide-I Spectroscopy of Wild Type and Mutant Monomers:} Mutation
at position B24 considerably influences the dimerization behaviour of
the hormone.\cite{desmond-jcpb-2019} Hence, the dynamics of the
hydrated F24A and F24G monomers was first considered. The infrared
lineshapes for residues along the dimerization interface and the same
selected -CO probes for the WT monomer are reported in Figure
S6. For the two mutant monomers (Figure
S6A for F24A and Figure
S6B for F24G) the frequency maximum for -CO at
position B24 is shifted from 1612.5 cm$^{-1}$ (WT) to 1614.5 cm$^{-1}$
(F24A) and 1628.5 cm$^{-1}$ (F24G), respectively. The amide-I band
maxima at positions B25 and B26 show differences for the the F24A
mutant but not for F24G and for position A19 the frequency maxima
shift to the blue (7 cm$^{-1}$) for F24A and to the red (6 cm$^{-1}$)
for F24G compared to WT. For all other -CO labels in the monomer the
differences between F24A and F24G are less than 14 cm$^{-1}$.\\

\begin{figure}[H]
\begin{center}
\includegraphics[width=0.99\textwidth]{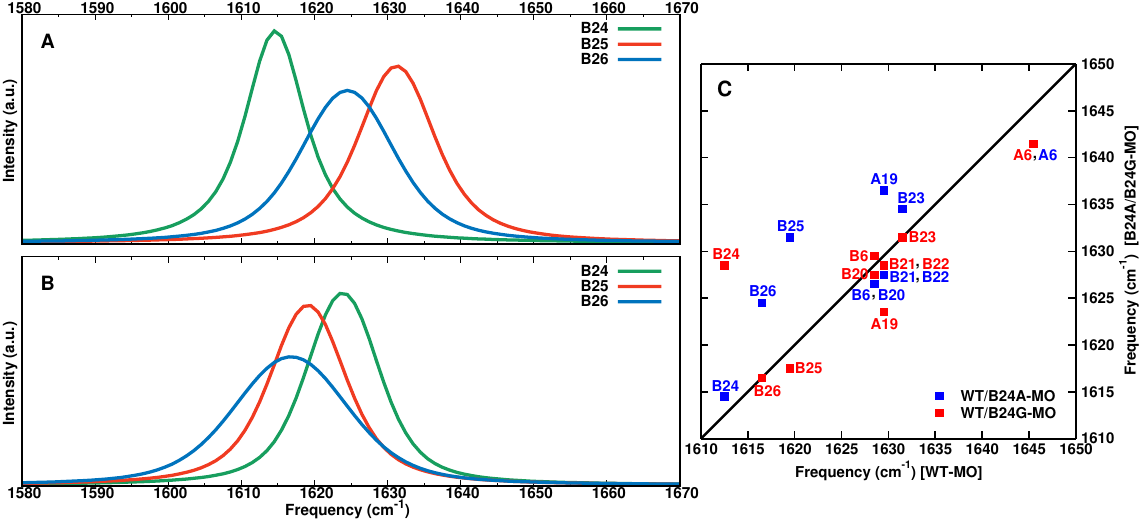}
\caption{1D IR spectra for monomeric mutants at position B24. Panels A
  and B report spectra for F24A (panel A) and F24G (panel B) for
  residues (B24-B26) at the dimerization interface, based on ``scan''
  for frequency calculations. Panel C compares the maximum frequency
  of the 1D IR spectra for selected residues (A6, A19, B6, B20-B26)
  between monomeric WT and mutants F24A and F24G.}
\label{fig:scan.momut}
\end{center}
\end{figure}

\noindent
A direct comparison of the maxima between the WT and the two mutant
monomers is given in Figure \ref{fig:scan.momut}C for selected -CO
probes, as for WT monomer and dimer (see Figure \ref{fig:scan.modi}).
The most pronounced differences in the maximum absorbances occur
around the mutation site whereas away from it they are minor, except
for -CO at position A19. Interestingly, residue TyrA19 is structurally
close to PheB24 (see Figure \ref{fig:structure}A) which explains the
dynamical coupling between the two sites that leads to a shift of
$\sim \pm 7$ cm$^{-1}$ and is also consistent with recent work on the
stability of B24-mutated insulin.\cite{MM.insulin:2018}\\

\begin{figure}[H]
\begin{center}
\includegraphics[width=0.99\textwidth]{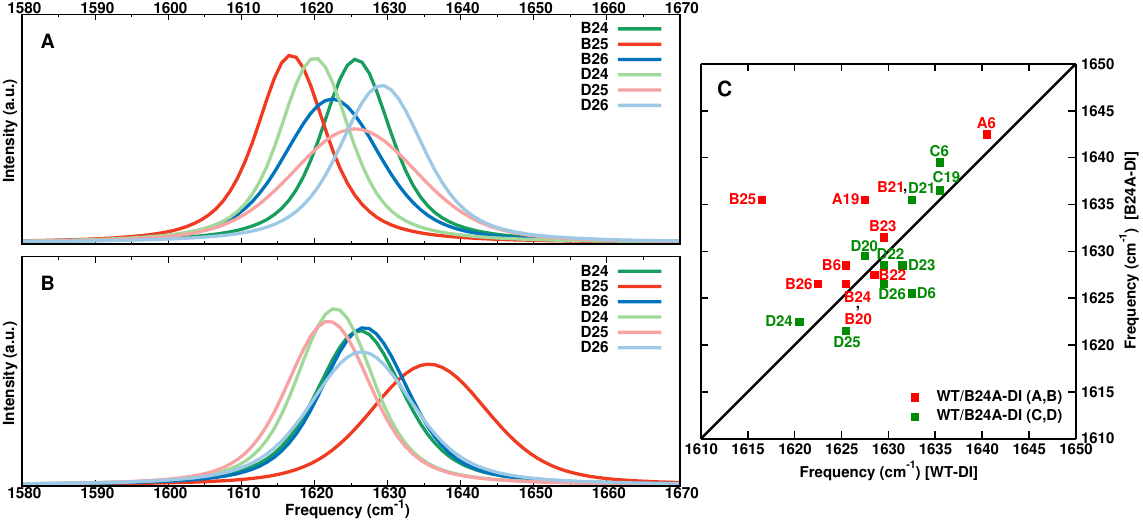}
\caption{1D IR spectra for WT (panel A) and the F24A (panel B) dimer
  for residues at the dimerization interface (B24-B26, D24-D26), based
  on ``scan'' for frequency calculation. Panel C compares the maximum
  frequency of the 1D IR spectra for selected residues (A6, A19, B6,
  B20-B26, C6, C19, D6, D20-D26) between the WT and F24A mutant
  dimer.}
\label{fig:scan.di24a}
\end{center}
\end{figure}

\noindent
{\it Amide-I Spectroscopy of Wild Type and Mutant Dimers:} The peak
frequencies for residues at the dimerization interface for the WT and
the F24A mutant are reported in Figures \ref{fig:scan.di24a}A and B
and directly compared for a larger number of residues, see Figure
\ref{fig:scan.di24a}C and S7. As for the
monomer, there are specific differences such as for TyrA19, PheB25,
and PheD25 which shift by up to 15 cm$^{-1}$ between the two
systems. For other residues the differences are considerably
smaller. For the F24G mutant differences persist, but are in general
smaller, see Figure S8. What is found from
simulations for both mutants is that residues are not necessarily
symetrically affected, in particular for those along the dimerization
interface. Also, depending on the modification at position B24 the
effects differ and may allow to distinguish between the different
insulin variants.

\subsection{Comparison of Amide-I Spectroscopy from Scan, Normal Mode and Map Analyses}
The three approaches to determine frequency trajectories considered
here (``scan'', ``INM'', and ``map'') differ considerably in terms of
computational expense and the formal approximations in applying
them. Scanning along the CO or amide-I normal mode for every snapshot
is computationally expensive as it requires for every snapshot to
carry out a 1-dimensional scan of the PES, representing it as a RKHS,
and solving the nuclear Schr\"odinger equation. As this needs to be
done for $\sim 10^5$ snapshots per nanosecond, such an approach does
not scale arbitrarily to larger systems and long time scales ($\mu$s
or longer). Compared to ``scan'', determining instantaneous normal
modes is computationally less demanding and the ``map'' approach is
also computationally efficient. In the following, the lineshapes from
the frequency trajectory for the WT monomer using the three methods
are compared.\\

\begin{figure}[H]
\begin{center}
\includegraphics[width=0.37\textwidth,angle=-90]{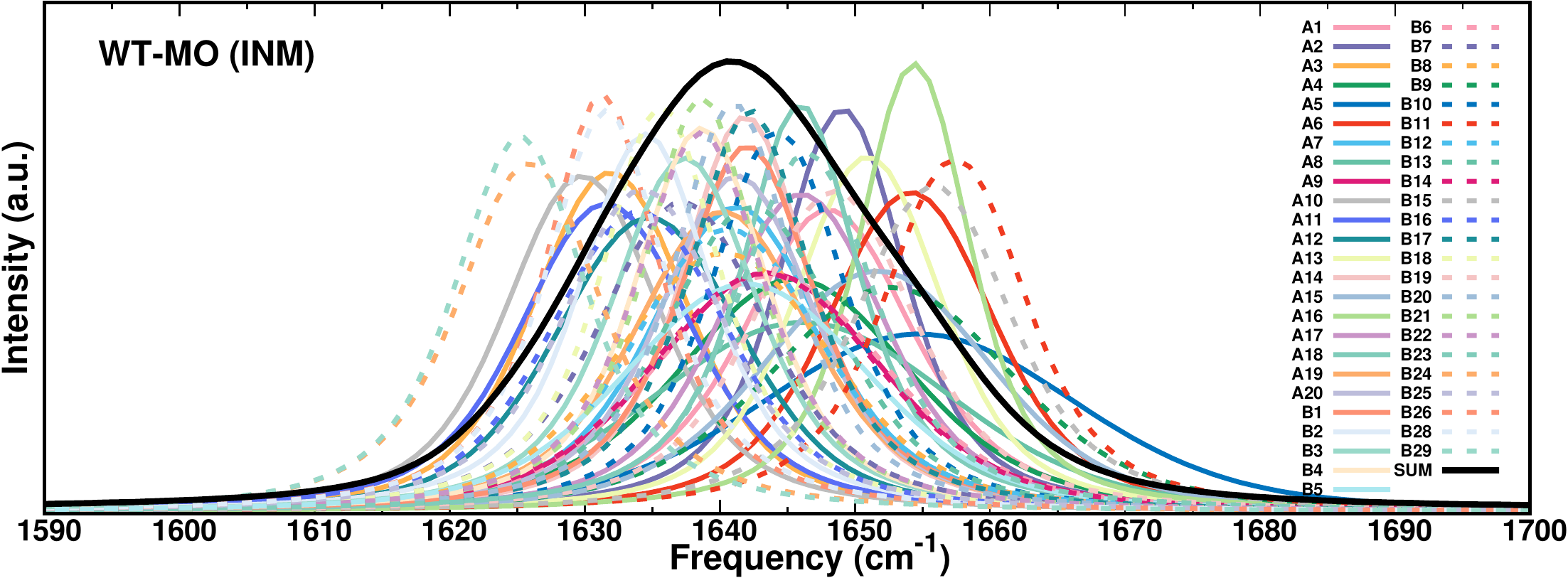}
\caption{1D IR spectra for all residues for the WT monomer from INM for the
  frequency calculations. The black line shows the superposition of
  all CO spectra compared with other single CO spectrum.}
\label{fig:INM.wtmo}
\end{center}
\end{figure}

\noindent
Figure \ref{fig:INM.wtmo} reports the 1d lineshapes for all residues
of the WT monomer from INM. As for ``scan'' the maxima of the
individual line shapes cover a range between 1625.5 cm$^{-1}$ and
1657.5 cm$^{-1}$ and the average spectra over all individual
lineshapes is centered at 1640.5 cm$^{-1}$ with a FWHM of 26
cm$^{-1}$, compared with 1630.5 cm$^{-1}$ and a FWHM of $\sim 30$
cm$^{-1}$ from ``scan'', see Figure \ref{fig:scan.monomer}. A direct
comparison of the frequency maxima for the WT monomer from ``scan''
and INM is reported in Figure S9A.\\
 
\begin{figure}[H]
\begin{center}
\includegraphics[width=0.37\textwidth,angle=-90]{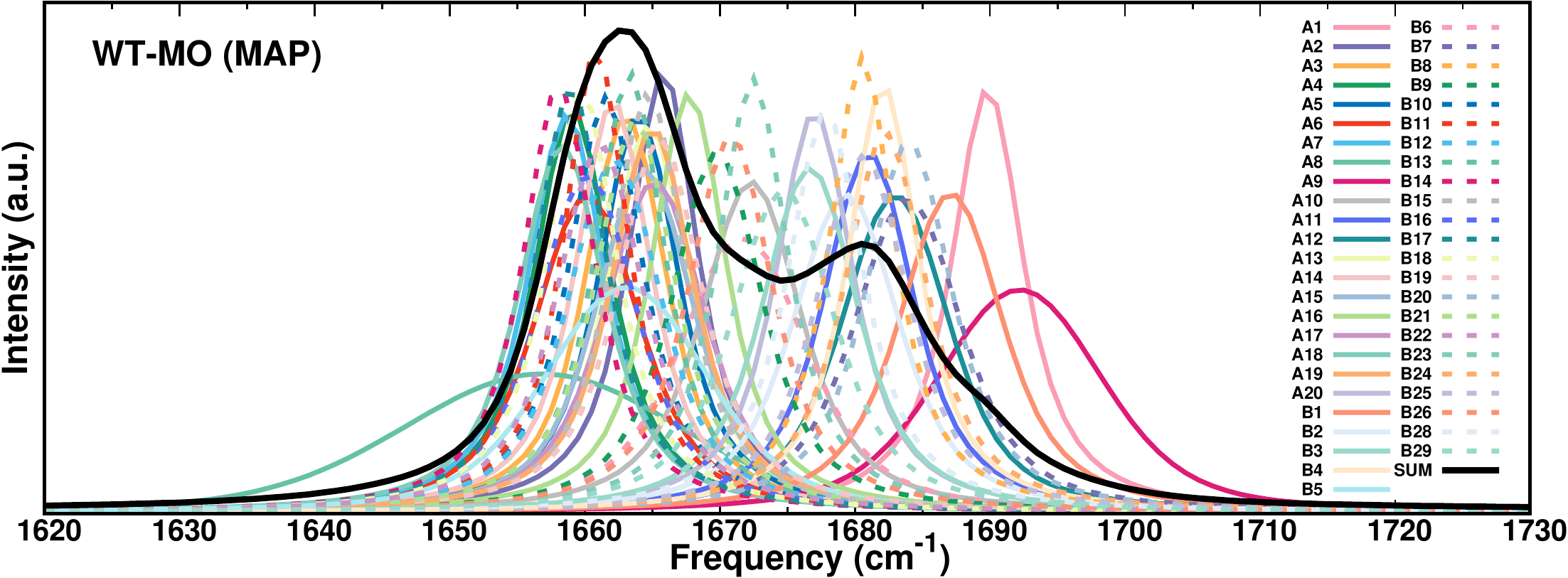}
\caption{1D IR spectra for all residues in WT monomer based on ``map'' for
  the frequency calculation. The labels for the individual line shapes
  are given in the panel and the overall sum is the solid black
  line. The line shapes are determined from 1 ns simulations and the
  snapshots analyzed are separated by 10 fs.}
\label{fig:map.wtmo}
\end{center}
\end{figure}

\noindent
The individual and total lineshapes from using the ``map'' frequencies
are reported in Figure \ref{fig:map.wtmo}. Again, the individual
frequency maxima span a range of $\sim 50$ cm$^{-1}$ and the FWHM
differ for the residues. Contrary to the overall line shape for the
monomer from ``scan'' and ``INM'', using the frequency map leads to an
infrared spectrum with two peaks. This shape is not consistent with
the experimentally observed IR
spectrum.\cite{tokmakoff:2016,tokmakoff:2016.2} Also, the frequency
maxima are somewhat displaced to higher frequencies and do not
correlate particularly well with the frequency maxima from ``scan''
(see Figure S9B). One possibility for these
differences may be the fact that for using ``map'' simulations with
constrained -CO are required. Also, the map used in the present work
was parametrized with respect to experiments and using a point
charge-based force field whereas the simulations in the present work
used multipoles.\\

\begin{figure}[H]
\begin{center}
\includegraphics[width=0.99\textwidth]{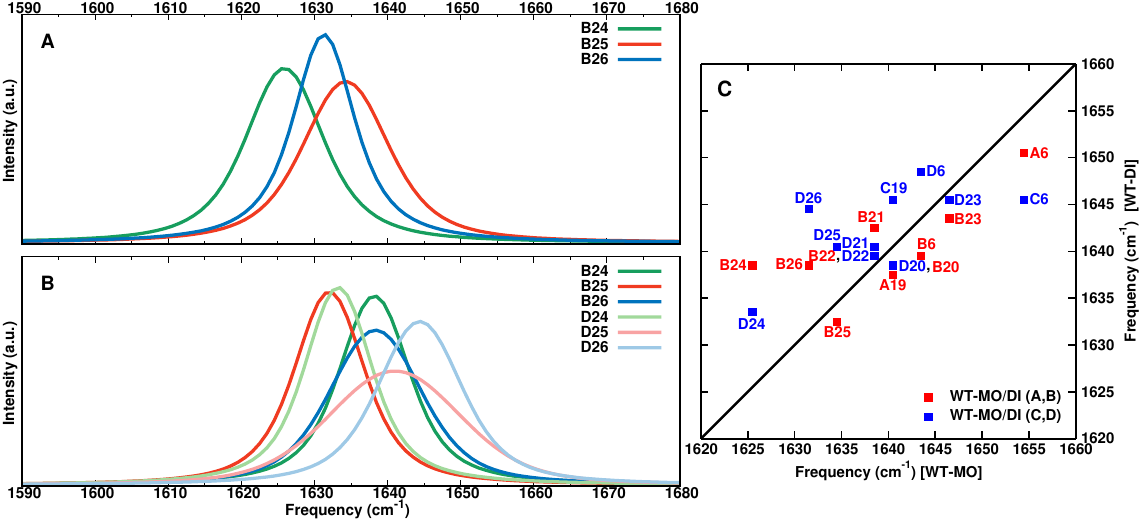}
\caption{1D IR spectra from INM for residues (B24-B26) and (B24-B26,
  D24-D26) at the dimerization interface for WT monomer (panel A) and
  WT dimer (panel B). Panel C compares the maximum frequency of the 1D
  IR spectra for the residues (A6, A19, B6, B20-B26, C6, C19, D6,
  D20-D26) between WT monomer and dimer.}
\label{fig:INM.modi}
\end{center}
\end{figure}

\begin{figure}[H]
\begin{center}
\includegraphics[width=0.99\textwidth]{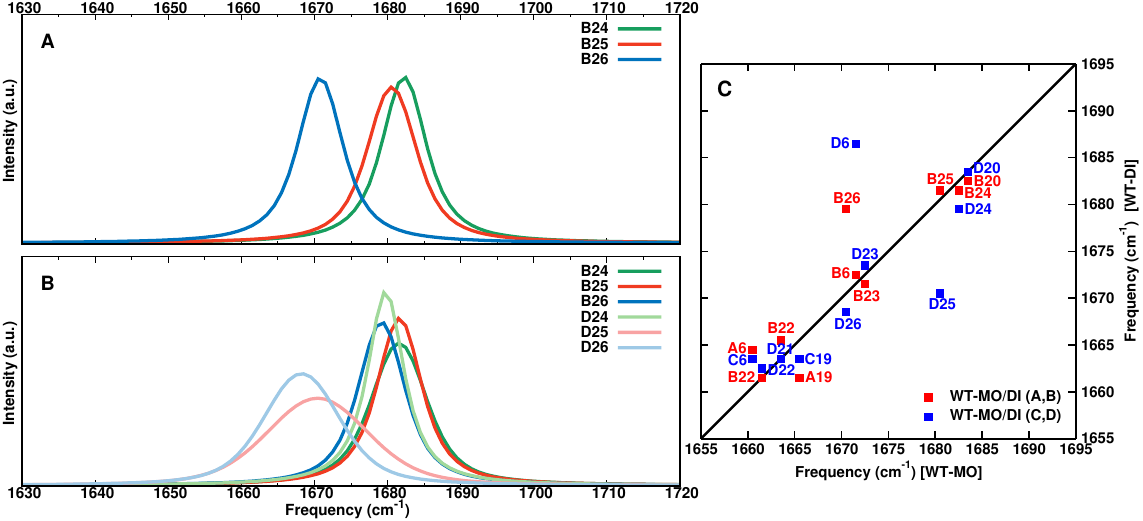}
\caption{1D IR spectra from "map" for residues (B24-B26) and (B24-B26,
  D24-D26) at the dimerization interface for WT monomer (panel A) and
  WT dimer (panel B). Panel C compares the maximum frequency of the 1D
  IR spectra for the residues (A6, A19, B6, B20-B26, C6, C19, D6,
  D20-D26) between WT monomer and dimer. The CO bond length is
  constrained in the MD simulations.}
\label{fig:map.constr.modi}
\end{center}
\end{figure}

\noindent
Next, the lineshapes for the residues involved in the dimerization
interface and the selection of other residues already considered until
now are analyzed for WT monomer and dimer for INM and ``map'', see
Figures \ref{fig:INM.modi}, \ref{fig:map.constr.modi},
S10, and S11. When using INM
it is again found that for the residues at the dimerization interface
the location of the frequency maxima in the two monomers differ and
also change compared with the isolated monomer (see Figure
\ref{fig:INM.modi}C). These effects are not only observed for residues
at the interface but also away from it.  Splitting for B/D24, B/D25,
and B/D26 are comparable or larger than with ``scan'' and blue/red
shifts are consistent for the two methods.\\

\noindent
For the analysis using ``map'' in Figure \ref{fig:map.constr.modi} it
is important to note that they do not use the same structures for
analysis as for ``scan'' and INM because the -CO bond lengths were
constrained. As for the other two methods the frequency maxima for B24
to B26 do not coincide for the monomer (Figure
\ref{fig:map.constr.modi}A) and the -CO labels in the two monomers
have their maxima at different frequencies in the dimer (Figure
\ref{fig:map.constr.modi}B). However, the actual frequency maxima
between the three methods differ. The effect of constrained and
flexible -CO in the MD simulations is reported in Figure S12. For a comparison of the maximum frequencies
for the three methods for B24 to B26 and D24 to D26 for direct
numerical comparison, see Table \ref{tab:comp}. Figure S13 reports a comparison of the map used here and an
alternative parametrization.\cite{skinner-map-JPCB2011} Consistent
with earlier work that compared the performance of different
maps,\cite{tokmakoff:2015} it is found that the two correlate quite
well (within a few cm$^{-1}$) except for residue B20 for which they
differ by $\sim 25$ cm$^{-1}$. It is noteworthy that for both,
scanning along the [CONH] normal mode (Figure S1) and for using ``map'' (Figure S9) compared with scanning along the CO
mode, the frequency maxima are shifted towards the blue, in accord
with experiment (frequency maximum $\sim 1650$
cm$^{-1}$).\cite{tokmakoff:2016,tokmakoff:2016.2}\\

\begin{table}
\begin{tabular}{|c||c|c|c|}
 Residue & Scan & INM & Map \\
 \hline 
   B24 & 1612.5&1625.5 &1682.5   \\
    \hline  
   B25 & 1619.5&1634.5 &1680.5    \\ 
    \hline 
   B26 & 1616.5&1631.5 &1670.5    \\
 \hline
\end{tabular}
\caption{Position of the frequency maxima of the 1D IR spectra
  for WT monomer using the three different approaches (``scan'',
  ``INM'', and ``map''). For ``scan'' and INM the CO probes are
  flexible while for "map" the structures were those from a simulation
  with constrained CO bond length.}
\label{tab:comp}
\end{table}

\noindent
Using ``map'' the labels at B/D25 and B/D26 show splittings comparable
to those from ``scan'' and INM whereas for B/D24 the splitting is only
1 to 3 cm$^{-1}$ which is considerably smaller than for the two other
methods. Nevertheless, the results from ``map'' also indicate that the
spectroscopic signatures of the residues at the dimerization interface
are not identical and differ from the monomer whereas for the other
residues considered the differences between monomer and dimer and the
two monomers within the dimer are smaller.\\

\noindent
In summary, all three methods agree in that a) the individual labels
have their frequency maxima at different frequencies and b) in
going from the WT monomer to the dimer the IR spectra of the labels
involved in dimerization split and shift. The magnitude of the
splitting and shifting differs between the methods which is not
surprising given their very different methodologies. For the two
mutants F24A and F24G the IR lineshapes using ``scan'' were determined
for the residues involved in the dimerization interface and a
selection of other residues, see Figure \ref{fig:structure}A. Compared
with the WT monomer and dimer, characteristic shifts were
found. \\

\subsection{Frequency Fluctuation Correlation Functions}
The frequency fluctuation correlation functions that can be computed
from the frequency time series contain valuable information about the
dynamics around a particular site considered, here the -CO groups of
every residue. Specifically, FFCFs were analyzed for labels along the
dimerization interface, for WT and the two mutant monomers and dimers,
from using frequencies determined from ``scan'' and INM. Before
discussing the FFCFs their convergence with simulation time is
considered as it has been observed that an extensive amount of data is
required.\cite{Salehi:N3-JPCB2019}\\

\noindent
For this, the first 1 ns and the entire 5 ns run for WT insulin
monomer was analyzed using ``scan''. For the 1 ns simulation snapshots
every 10 fs and every 2 fs were analyzed (see Figure S14 top and middle row) and every 10 fs for the 5 ns
simulations (Figure S14 bottom row). The
computational resources required for such an analysis are
considerable. Using 8 processors, the analysis of the 1 ns simulation
for $10^5$ snapshots (saved every 10 fs) takes 400 hours for a single
spectroscopic probe. Figure S14 shows that except
for one feature at $\sim 3$ ps for residue B26 the FFCFs from the 1 ns
simulation with saving every 10 fs and every 2 fs are very similar. On
the contrary, using snapshots from the 5 ns simulation leads to
reducing the fluctuations in the FFCFs and determinants such as the
static component (the value at a correlation time of 4 ps) are higher
from the longer simulation. A quantitative comparison for the time
scales, amplitudes and static component (see Eq. \ref{eq:ffcf}) is
provided in Figure S15 and in Table S2. The amplitudes and short decay times of all
fits are within a few percent. The picosecond time scale ($\tau_2$)
can differ by up to 30 \% (B26) and the offset $\Delta_0$ can differ
by a factor of two or more. To balance computational expense and
quality of data, the remaining analysis was carried out with data from
the 1 ns simulation with snapshots recorded every 10 fs.\\

\begin{figure}[H]
\begin{center}
\includegraphics[width=0.99\textwidth]{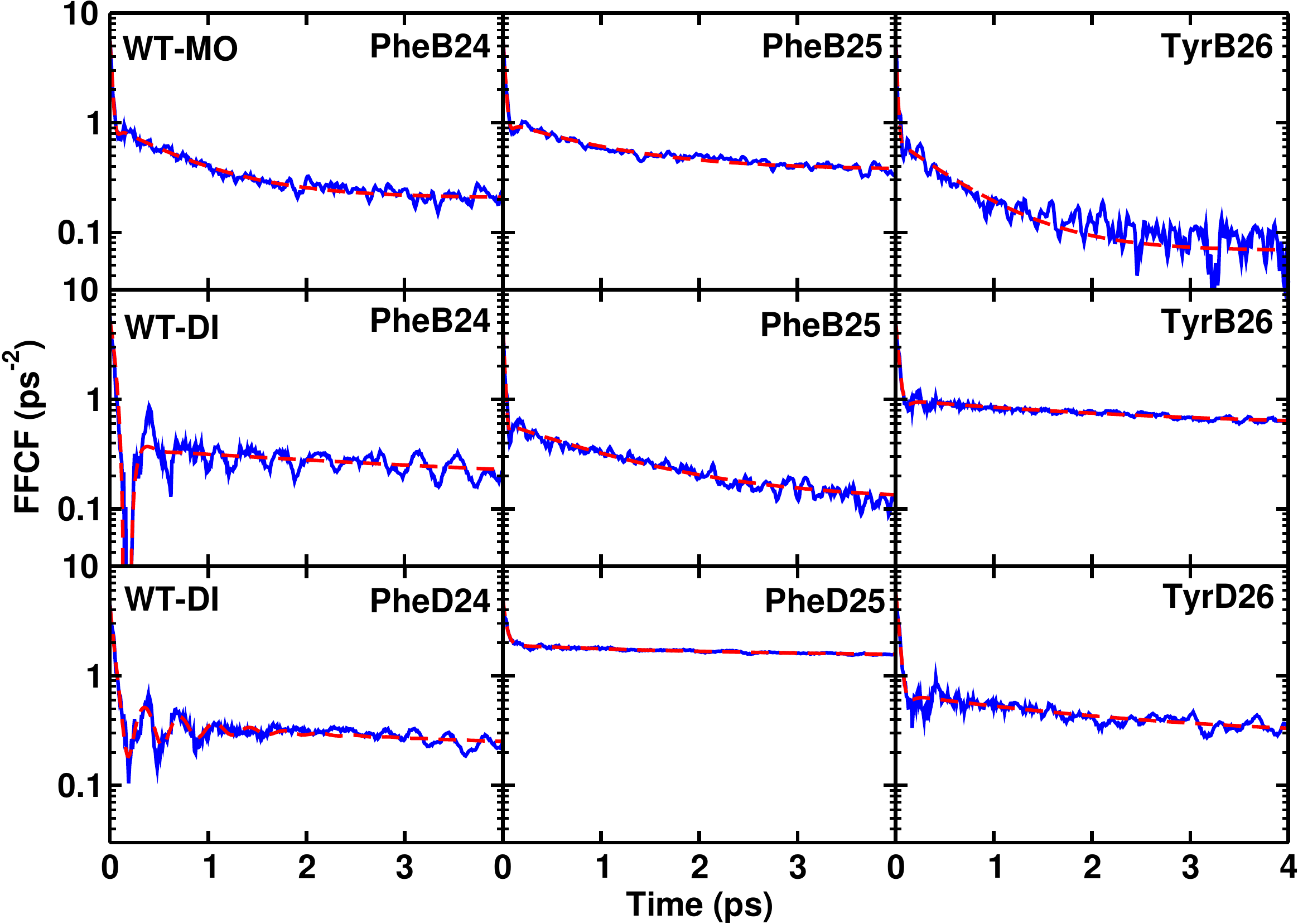}
\caption{Comparison of the FFCFs for WT monomer and dimer for residues
  B24 to B26 at the dimerization interface. The frequencies are based
  on ``scan'' and snapshots from the 1 ns simulation, saved every 10
  fs were analyzed.}
\label{fig:ffcf.scan}
\end{center}
\end{figure}

\noindent
The FFCFs for B24 to B26 of the insulin monomer and the two monomers
within the dimer are reported in Figure \ref{fig:ffcf.scan} together
with the fits to Eq. \ref{eq:ffcf}. For the three labels from the
monomer simulations the FFCFs differ in the longest decay time and the
offset $\Delta_0$. As for the infrared spectra, the three -CO labels
exhibit different environmental dynamics. When compared with the two
monomers in the insulin dimer these differences are even more
pronounced. In general, all decay times increase to between 1 ps and
$\sim 13$ ps and the offset can be up to 5 times larger than for the
monomer. This is owed to the considerably restrained dynamics of the
residues at the dimerization interface compared with the free
monomer.\\

\begin{table}
\begin{tabular}{c|rrr|rr|r|rr}
&$a_{1}$ & $\gamma$ &$\tau_{1}$ &$a_{2}$ &$\tau_{2}$ &$\Delta_{0}$ &$a_{3}$ &$\tau_{3}$\\
\hline
\textbf{WT monomer}\\
B24 &4.64 &25.44 &0.025 & 0.75&0.74 &0.21&&\\
B25 &4.94 &22.19 &0.028 & 0.66&0.98  &0.37\\
B26 &4.97 &21.82 &0.019 & 0.62&0.62  &0.07\\
\hline
\textbf{WT dimer M1} & & & & & & \\
B24 &4.80 &14.50 &0.080 &0.23 &4.72  &0.13  \\
B25 &3.92 &27.74 &0.023 &0.49 &1.15   &0.12\\
B26 &4.12 &16.36 &0.038 &0.44 &2.51   &0.55 \\
\hline
\textbf{WT dimer M2} & & & & & & \\
D24 &0.30 &17.59 &0.56 &3.68 &0.039  &0.18 &0.19&4.08\\
D25 &3.17 & &0.033 &0.41 &2.32  &1.50 \\
D26 &5.32 &13.49 &0.040 &0.42 &2.10  &0.27  \\
\hline
\hline
\textbf{F24A monomer} & & & & & & \\
B24 &4.94 &29.51 &0.020 &0.51 &0.61  &0.07\\
B25 &4.37 &13.45 &0.020 &0.64 &0.79  &0.21 \\
B26 &4.11 &25.05 &0.027 &0.63 &1.38  &0.45 \\
\hline
\textbf{F24A dimer M1} & & & & & & \\
B24 &4.90 &14.61 &0.046 &0.33 &1.68  & 0.41 \\
B25 &3.19 & &0.028 &0.62 &1.81  &1.08 \\
B26 &2.15 & &0.040 &0.34 &1.89  &0.48\\
\hline
\textbf{F24A dimer M2} & & & & & & \\
D24 & 1.27 & &0.043 &0.31 &1.17  &0.36 \\
D25 & 1.39 & &0.031 &0.32 &1.10  &0.51 \\
D26 & 4.91 &13.72 &0.039 &0.60 &1.40  &0.63 \\
\hline
\hline
\textbf{F24G monomer} & & & & & & \\
B24 &4.72 &29.74 &0.032 &0.43 &1.24 &0.26\\
B25 &4.57 &16.46 &0.019 &0.58 &0.81 &0.24  \\
B26 &4.60 &25.73 &0.022 &0.59 &0.54 &0.04 &0.51&7.89  \\
\hline
\textbf{F24G dimer M1} & & & & & & \\
B24 &1.42 & &0.028 &0.21 &1.02  &0.37   \\
B25 &3.70 & &0.018 &0.48 &1.18  &0.24 \\
B26 &3.87 & &0.029 &0.68 &1.90  &0.88 \\
\hline
\textbf{F24G dimer M2} & & & & & & \\
D24 &2.50 &38.76 &0.016 &0.30 &1.29  &0.17  \\
D25 &1.53 & &0.030 &0.25 &1.70  &0.65  \\
D26 &3.94 &5.32 &0.042 &0.27 &2.14  &0.23 \\
\end{tabular}
\caption{Parameters from fitting the FFCF to Eq. \ref{eq:ffcf} for
  frequencies from ``scan'' for the selected residues (B24-B26 and
  D24-D26). The amplitudes $a_1$ to $a_3$ in ps$^{-2}$, the decay
  times $\tau_1$ to $\tau_3$ in ps, the parameter $\gamma$ in
  ps$^{-1}$, and the offset $\Delta_0$ in ps$^{-2}$. For residues D24
  in monomer M2 from the WT dimer and B26 in the F24G monomer the
  third time scale is required for a good fit.}
\label{tab:ffcf.fit}
\end{table}

\noindent
Comparing the two monomer mutants with the WT it is found that the
picosecond component is comparable whereas $\Delta_0$ is similar (for
F24A) or somewhat larger (for F24G), see Table
\ref{tab:ffcf.fit}. When moving to the mutant dimers, the differences
with their monomeric counterparts are considerably smaller than for
the WT system. This is likely to be related to a weakening of the F24A
and F24G dimers which also allows water to penetrate more or less
deeply into the dimer interface.\cite{MM.insulin:2018} Overall, the
dynamics still is slowed down in the mutant dimers by up to a factor
of two compared with the mutant monomer but the effects are
considerably less pronounced than for the WT systems.\\

\noindent
FFCFs were also determined from frequency trajectories determined from
the INMs for the three residues at the dimerization interface, see
Table S3. The findings are similar to those from
analyzing frequencies from ``scan'' whereas the actual numerical
values for amplitudes, decay times and offset differ somewhat.\\

\section{Conclusion}
The present work demonstrates that WT insulin monomer and dimer and
mutant monomers and mutant dimers lead to different spectroscopic and
dynamical signatures for residues along the dimerization
interface. This is found - to different extent - for all three
approaches used for computing the frequency trajectory (``scan'', INM,
``map'') and suggests that the overall findings do not depend strongly
on the way how these frequencies are determined. The center frequency
and FWHM for insulin monomer are in qualitative (scan along CO INM) or
even quantitative (scan along [CONH] INM) agreement with experiment
which, together with earlier investigations of the spectroscopy and
dynamics of and around
NMA,\cite{NMA2dir-MM-2014,NMA-MM-2015,MM.rev.jcp:2020} provide a
validation of the computational model. It is noteworthy that using one
single parametrization for the -CO stretch and the multipoles on the
[CONH] moiety of the peptide bond the experimentally observed FWHM for
the protein is correctly described.\\

\noindent
The fact that the stability differences between WT and mutant (here at
position B24)\cite{MM.insulin:2018} insulin dimer are also reflected
in the spectroscopy and dynamics of WT and mutant insulin monomers and
dimers suggests that spectroscopic investigations can be used to
provide information about the association thermodynamics. This follows
earlier suggestions for characterizing protein-ligand
binding\cite{Suydam.halr.sci.2006} which are supported by atomistic
simulations.\cite{vibvtark:PCCP2017} For insulin this is particularly
relevant because except for the WT dimer direct thermodynamic
information about its stability appears to be missing. Replacing a
thermodynamic approach by a spectroscopic characterization is an
attractive alternative. The present work suggests that by combining
quantitative simulations with modern experiments is a potentially
useful way to obtain pharmacologically relevant information such as
the strength of the modified insulin dimers.\\

\section*{Supporting Information}
The supporting information reports further comparison of the infrared
spectra for WT and mutant insulin monomer and dimer. Additional
validation of the FFCF and comparisons of two different spectroscopic
maps are provided as well.

%\section*{Data Availability Statement}

\section*{Acknowledgments}
This work was supported by the Swiss National Science Foundation
grants 200021-117810, 200020-188724, the NCCR MUST, and the University
of Basel which is gratefully acknowledged. The authors thank
Profs. T. la Cour Jansen and A. Tokmakoff for valuable correspondence.

\bibliography{ins}

\begin{figure*}
\centering \includegraphics[width=0.8\textwidth]{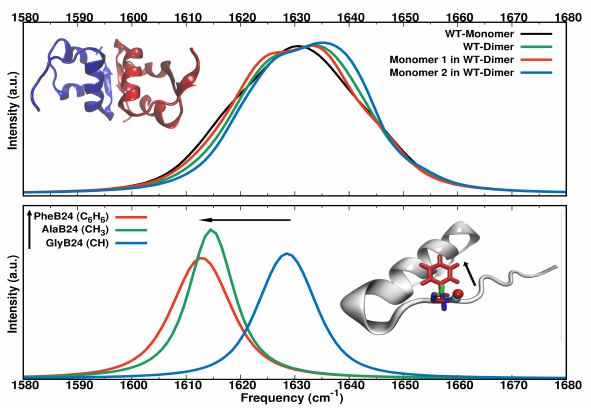}
\captionsetup{labelformat=empty}
\caption{TOC Graphic}
\end{figure*}

\end{document}

\section{Additional Lineshapes}

\begin{figure}[H]
\begin{center}
\includegraphics[width=0.99\textwidth]{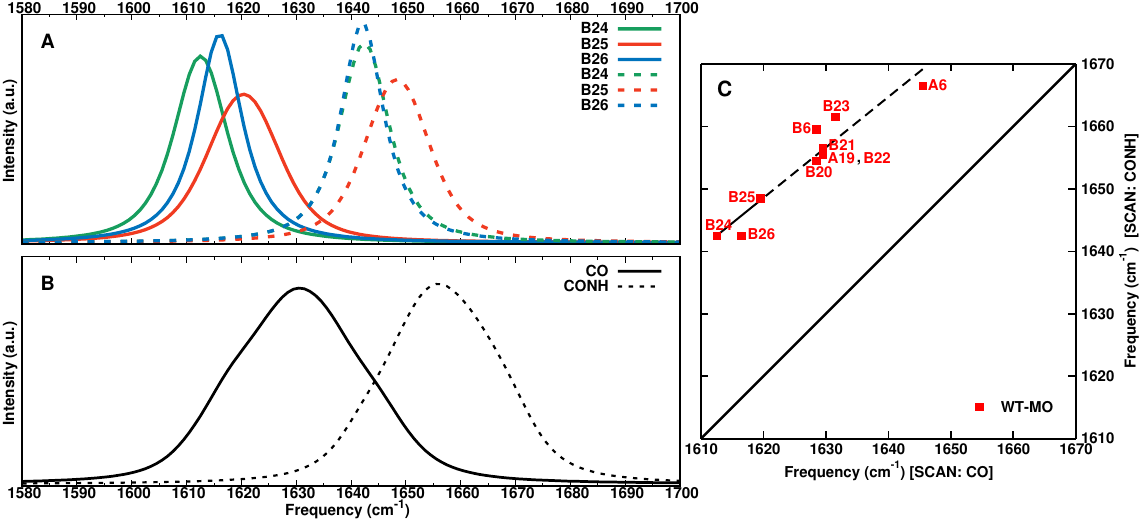}
\caption{Comparison for
  scanning along the CO (solid line) and CONH (amide-I, dashed line)
  normal modes for "scan" for the insulin monomer. Panel A: 1D IR
  spectra for residues (B24-B26), panel B: the sum frequency of all
  the residues and panel C: Comparison of the maximum frequency of the
  1D IR spectra for the selected residues (A6, A19, B6, B20-B26). The black dashed line shows the linear regression  with regression coefficient (slope) of 0.81 and correlation coefficient of 0.95. The
  analysis is done for 1 ns simulation and the snapshots analyzed are
  separated by 10 fs.  The frequency maxima from scanning along the
  [CONH] INM are shifted to the blue, in accord with the experimental
  observations.\cite{tokmakoff:2016,tokmakoff:2016.2}}
\label{sifig:scan.co.conh}
\end{center}
\end{figure}

\begin{figure}[H]
\begin{center}
\includegraphics[width=0.37\textwidth,angle=-90]{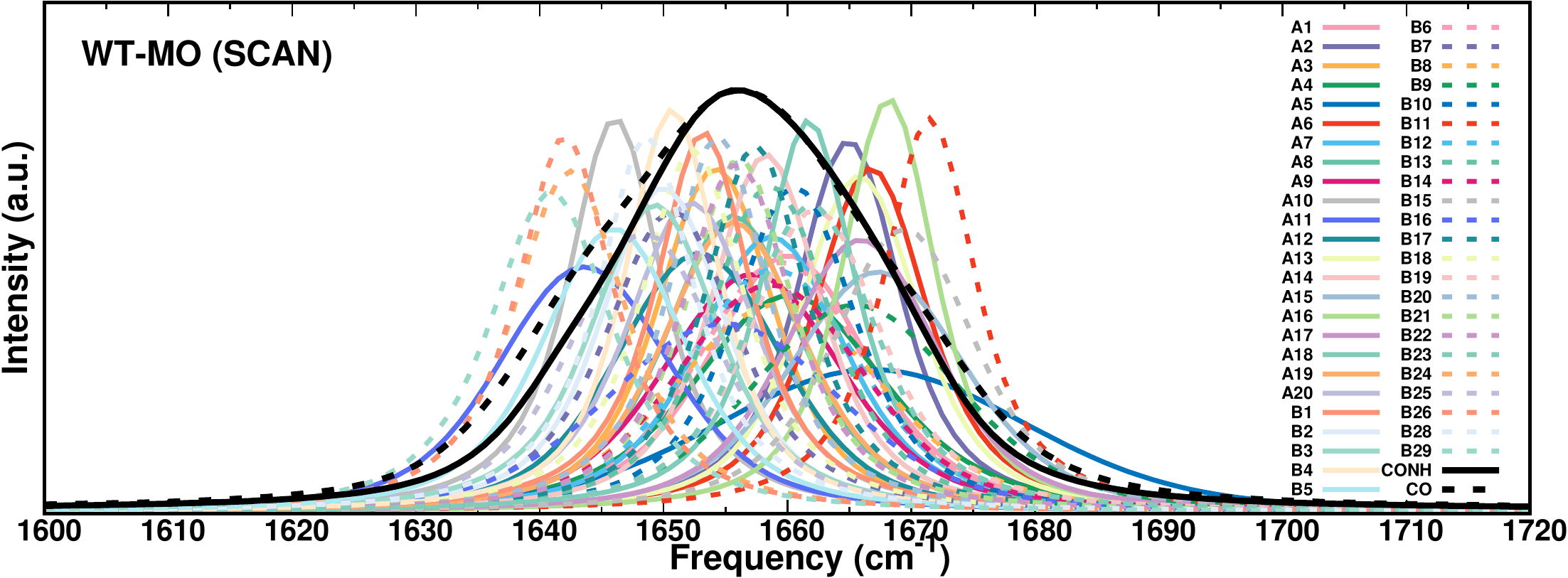}
\caption{1D IR spectra for all residues in WT monomer based on
  ``scan'' for frequency calculations along the amide-I normal
  mode. The labels for the individual line shapes are given in the
  panel and the overall sum is the solid black line. The line shapes
  are determined from 1 ns simulations and the snapshots analyzed are
  separated by 10 fs.}
\label{sifig:scan.conh}
\end{center}
\end{figure}

\begin{figure}[H]
\begin{center}
\includegraphics[width=0.99\textwidth]{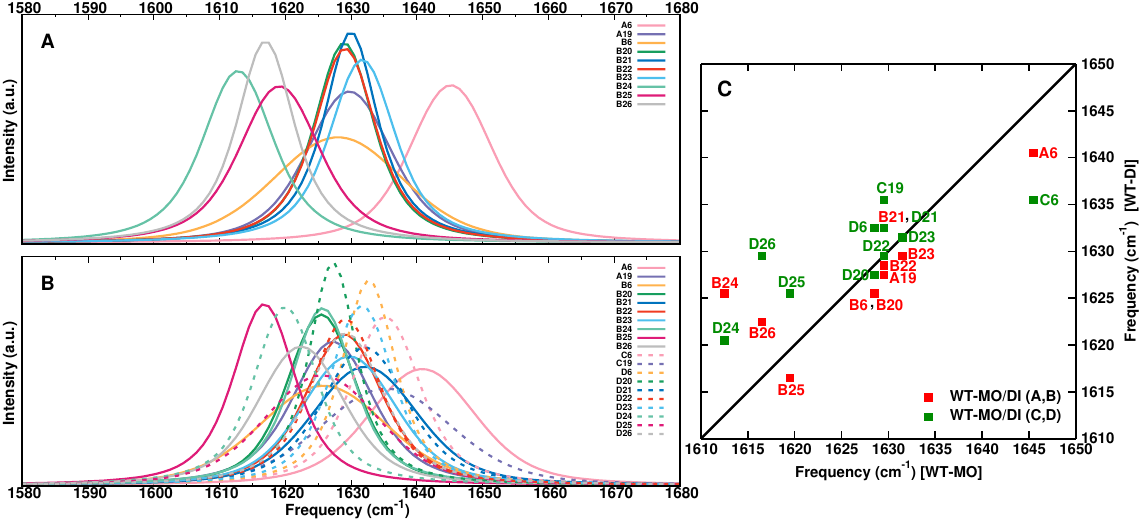}
\caption{1D IR spectra for WT monomer (panel A) and dimer (panel B)
  for selected residues (A6, A19, B6, B20-B26) and (A6, A19, B6,
  B20-B26, C6, C19, D6, D20-D26), using ``scan'' for the frequency
  calculation. Panel C: Comparison for the maximum frequency of the 1D
  IR spectra for the selected residues between WT monomer and dimer.}
\label{sifig:scan.modiall}
\end{center}
\end{figure}

\begin{figure}[H]
\begin{center}
\includegraphics[width=0.37\textwidth,angle=-90]{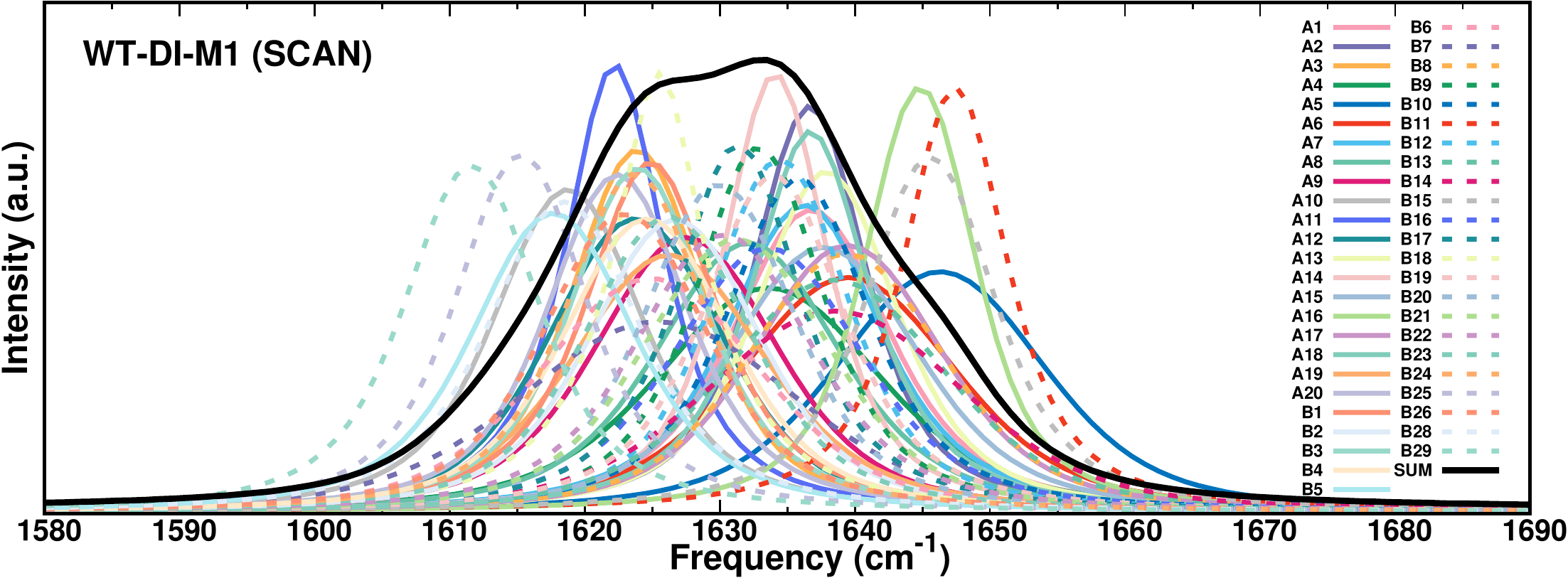}
\caption{1D IR spectra for all residues of monomer 1 (M1) in the WT
  dimer based on ``scan'' for the frequency calculation along the CO
  normal mode. The labels for the individual line shapes are given in
  the panel and the overall sum is the solid black line. The line
  shapes are determined from 1 ns simulations and the snapshots
  analyzed are separated by 10 fs.}
\label{sifig:scan.dim.mo1}
\end{center}
\end{figure}

\begin{figure}[H]
\begin{center}
\includegraphics[width=0.37\textwidth,angle=-90]{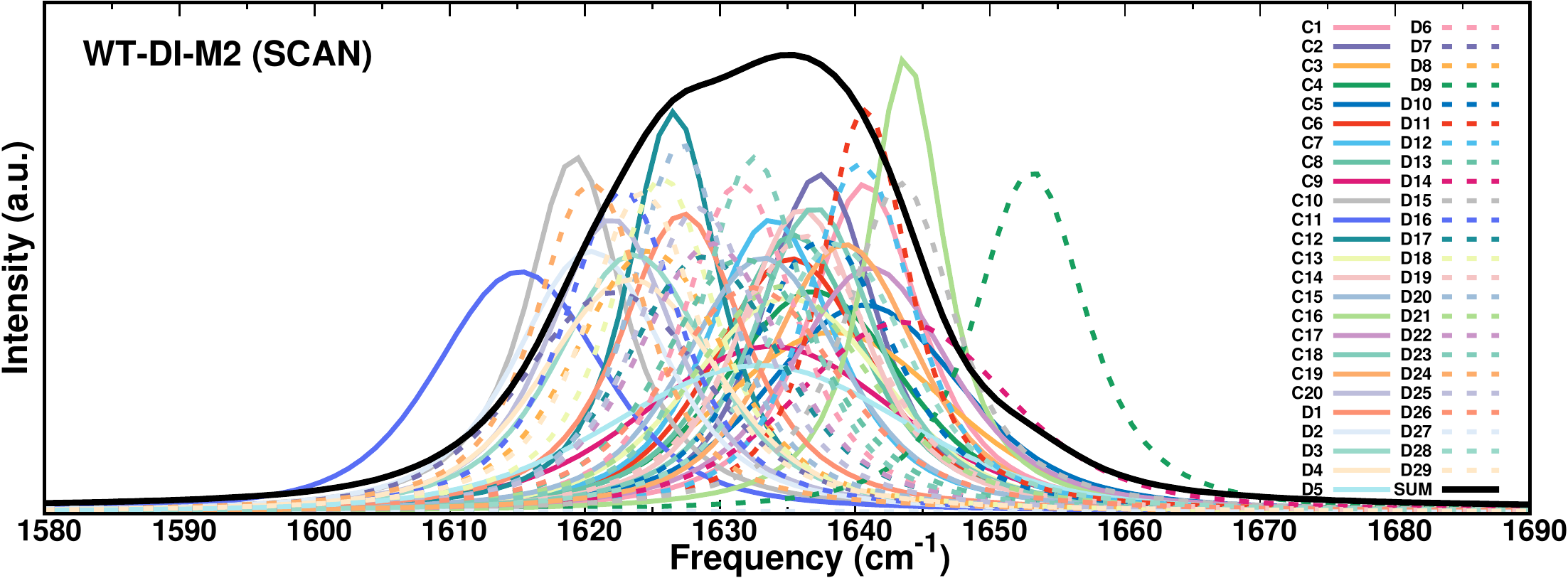}
\caption{1D IR spectra for all residues of monomer 2 (M2) in the WT
  dimer based on ``scan'' for the frequency calculation along the CO
  normal mode. The labels for the individual line shapes are given in
  the panel and the overall sum is the solid black line. The line
  shapes are determined from 1 ns simulations and the snapshots
  analyzed are separated by 10 fs.}
\label{sifig:scan.dim.mo2}
\end{center}
\end{figure}

\begin{figure}[H]
\begin{center}
\includegraphics[width=0.99\textwidth]{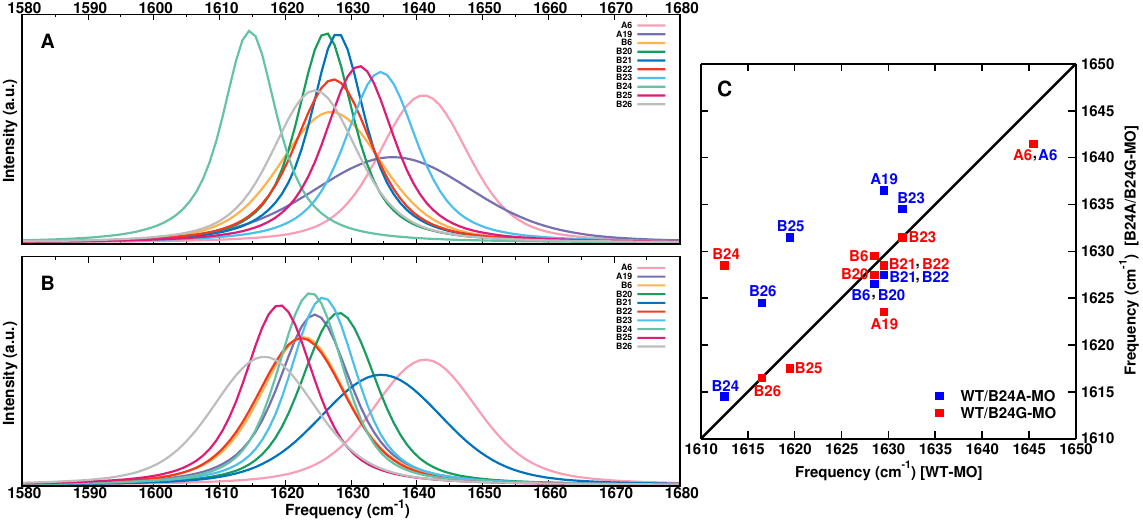}
\caption{1D IR spectra for the F24A (panel A) and F24G (panel B)
  mutant monomers for selected residues (A6, A19, B6, B20-B26) using
  ``scan''. Panel C: Comparison for the maximum frequency of the 1D IR
  spectra for the selected residues between WT and F24A/F24G
  monomers.}
\label{sifig:scan.momut.all}
\end{center}
\end{figure}

\begin{figure}[H]
\begin{center}
\includegraphics[width=0.99\textwidth]{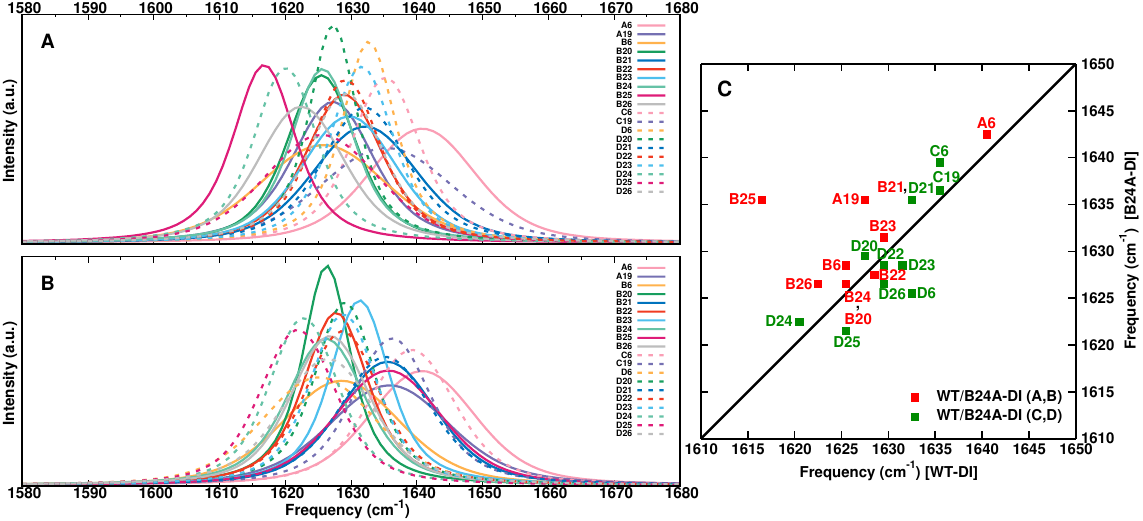}
\caption{1D IR spectra for the WT (panel A) and F24A (panel B) mutant
  dimers for selected residues (A6, A19, B6, B20-B26, C6, C19, D6,
  D20-D26) based on ``scan'' for the frequency calculations. Panel C:
  Comparison between maximum frequency of 1D IR spectra for the
  selected residues between WT and F24A mutant dimers.}
\label{sifig:scan.di24aall}
\end{center}
\end{figure}

\begin{figure}[H]
\begin{center}
\includegraphics[width=0.99\textwidth]{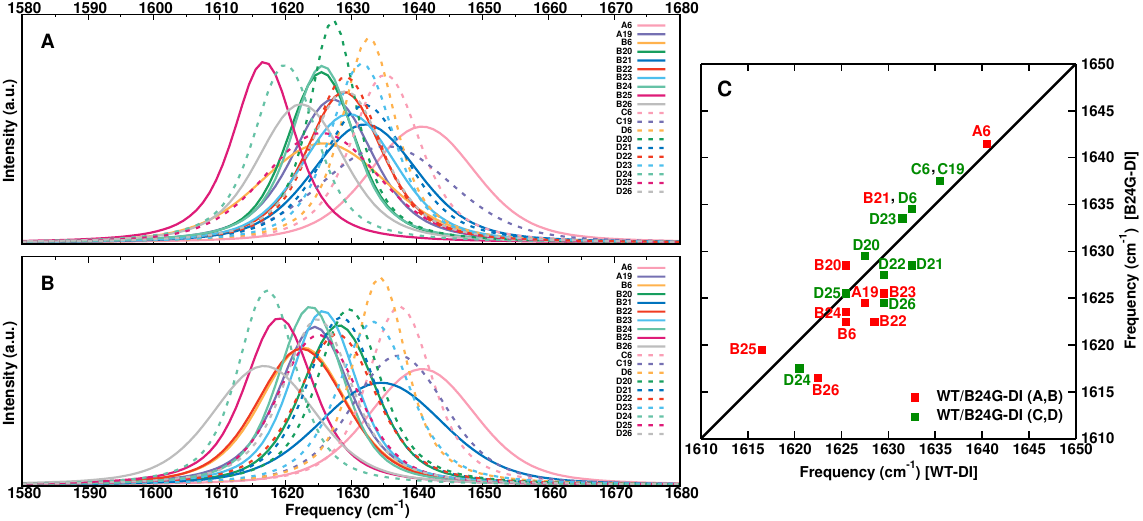}
\caption{1D IR spectra for the WT (panel A) and F24G (panel B) mutant
  dimers for selected residues (A6, A19, B6, B20-B26, C6, C19, D6,
  D20-D26) based on ``scan'' for the frequency calculations. Panel C:
  Comparison between maximum frequency of 1D IR spectra for the
  selected residues between WT and F24G mutant dimers.}
\label{sifig:scan.di24gall}
\end{center}
\end{figure}

\begin{figure}[H]
\begin{center}
\includegraphics[width=0.8\textwidth]{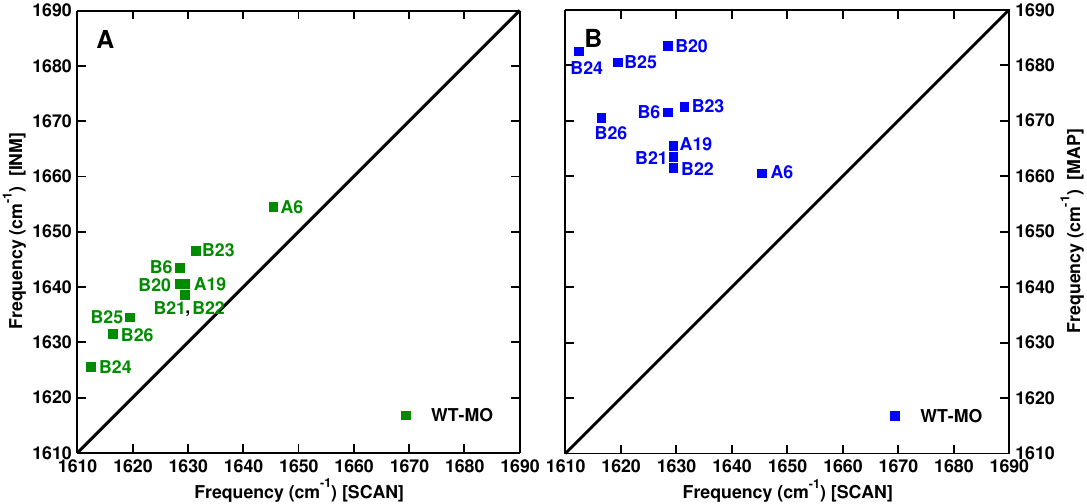}
\caption{Comparison of the maximum frequency of the 1D IR spectra
  between ``scan'' and ``INM'' (panel A) and "scan" and ``map'' (panel
  B) for the selected residues (A6, A19, B6, B20-B26, C6, C19, D6,
  D20-D26) for WT monomer. The CO probes are flexible in the
  simulations analyzed with ``scan'' and ``INM'' and constrained for
  the one using ``map''.}
\label{sifig:comp.scan.INM.map}
\end{center}
\end{figure}

\begin{figure}[H]
\begin{center}
\includegraphics[width=0.99\textwidth]{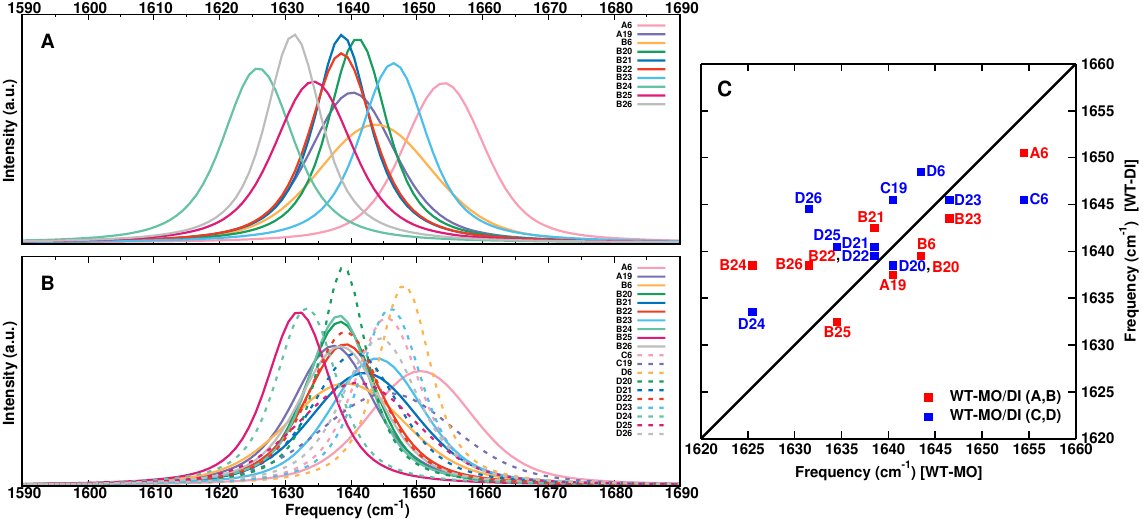}
\caption{1D IR spectra from INM. Panel A: WT monomer and panel B: WT
  dimer for selected residues (A6, A19, B6, B20-B26) and (A6, A19, B6,
  B20-B26, C6, C19, D6, D20-D26), respectively. Panel C: Comparison
  between maximum frequency of 1D IR spectra for the selected residues
  between monomer and dimer.}
\label{sifig:INM.modiall}
\end{center}
\end{figure}

\begin{figure}[H]
\begin{center}
\includegraphics[width=0.99\textwidth]{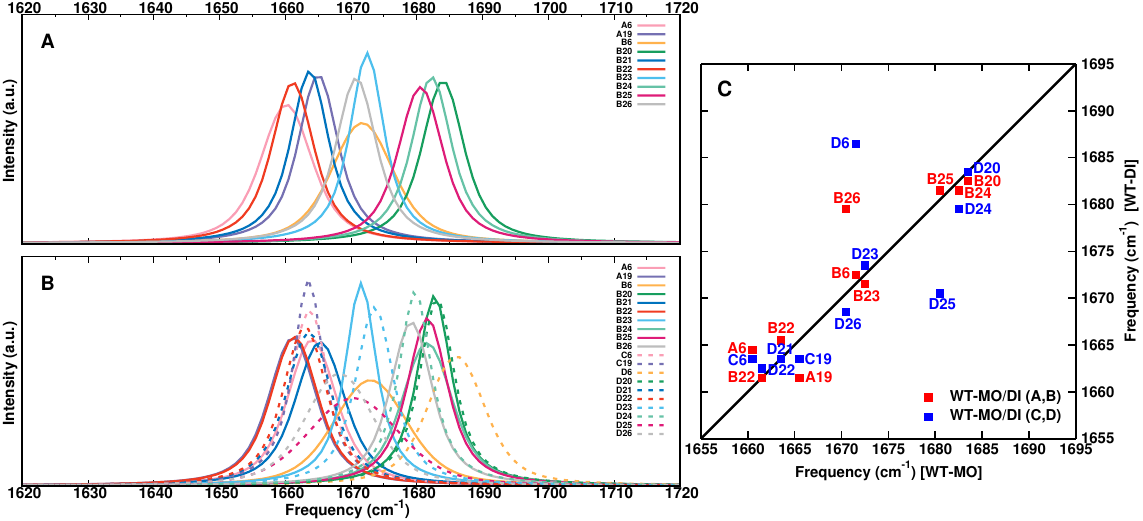}
\caption{1D IR spectra from "map" from simulations with constrained CO
  bond length. Panel A: WT monomer and panel B: WT dimer for selected
  residues (A6, A19, B6, B20-B26) and (A6, A19, B6, B20-B26, C6, C19,
  D6, D20-D26), respectively. Panel C: Comparison between maximum
  frequency of 1D IR spectra for the selected residues between monomer
  and dimer.}
\label{sifig:map.modiall}
\end{center}
\end{figure}

\begin{figure}[H]
\begin{center}
\includegraphics[width=0.99\textwidth]{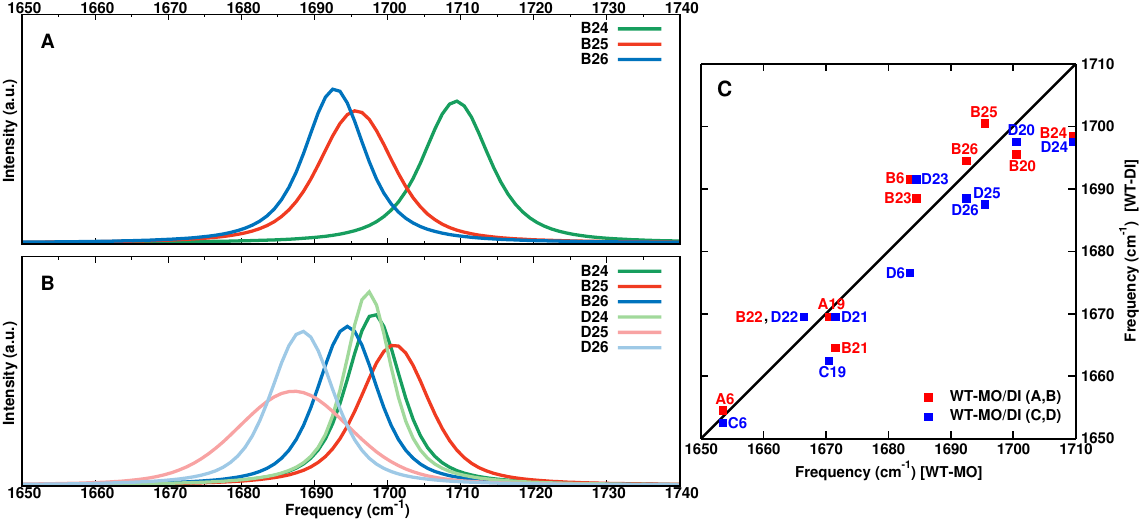}
\caption{1D IR spectra from "map" from simulations with flexible CO
  probe. Panel A: WT monomer and panel B: WT dimer for residues at the
  dimerization interface (B24-B26) and (B24-B26, D24-D26),
  respectively. Panel C: Comparison between maximum frequency of 1D IR
  spectra for residues (A6, A19, B6, B20-B26) and (A6, A19, B6,
  B20-B26, C6, C19, D6, D20-D26), respectively between WT monomer and
  dimer.}
\label{fig:map.flex.modi}
\end{center}
\end{figure}

\section{Comparison between two Different Maps}
As a separate test, a different map\cite{skinner-map-JPCB2011} is used
in which the frequency shift due to the dihedral angles $(\phi, \psi)$
between neighboring peptide units are included. Here the map
parametrization is
\begin{equation}\label{eq:4}
    \omega_i=1684+7729E_{C_{i}}+3576E_{N_{i}}
\end{equation}
and the local frequency is
\begin{equation}\label{eq:6}
    \omega_i^b=\omega_i+\Delta\omega_N(\phi_{i-1},\psi_{i-1})+\Delta\omega_C(\phi_{i+1},\psi_{i+1})
\end{equation}
Based on the $(\phi, \psi)$ angles for $i$th chromophore,
$\Delta\omega_{N}$ and $\Delta\omega_{C}$ are the contributions from
($i-1$)th and ($i+1$)th residues.

\begin{table}
\begin{tabular}{|c||c|c|}
 \hline\hline
 \multicolumn{3}{|c|}{Map Frequencies} \\
 \hline\hline
 Residue & Skinner Map & Tokmakoff Map \\
 \hline 
    A6 &  1654.50  &    1660.50 \\
     \hline 
   A19 &  1668.50  &    1665.50 \\
     \hline 
    B6 &  1671.50  &    1671.50\\
     \hline 
   B20 &  1715.50  &    1683.50\\
  \hline 
   B21 &  1660.50  &    1663.50\\
    \hline 
   B22 &  1666.50  &    1661.50\\
    \hline 
   B23 &  1677.50  &    1672.50\\
    \hline 
   B24 &  1676.50  &    1682.50\\
    \hline 
   B25 &  1692.50  &    1680.50 \\ 
    \hline 
   B26 &  1656.50  &    1670.50\\
 \hline
 \hline
\end{tabular}
\caption{Position of the frequency maxima from applying two maps
  (Skinner\cite{skinner-map-JPCB2011} and
  Tokmakoff\cite{Tokmakoff-map-jcp-2013}) to the same trajectory for
  WT insulin monomer for selected residues.}
\label{sitab:map}
\end{table}

\begin{figure}[H]
\begin{center}
\includegraphics[width=0.45\textwidth]{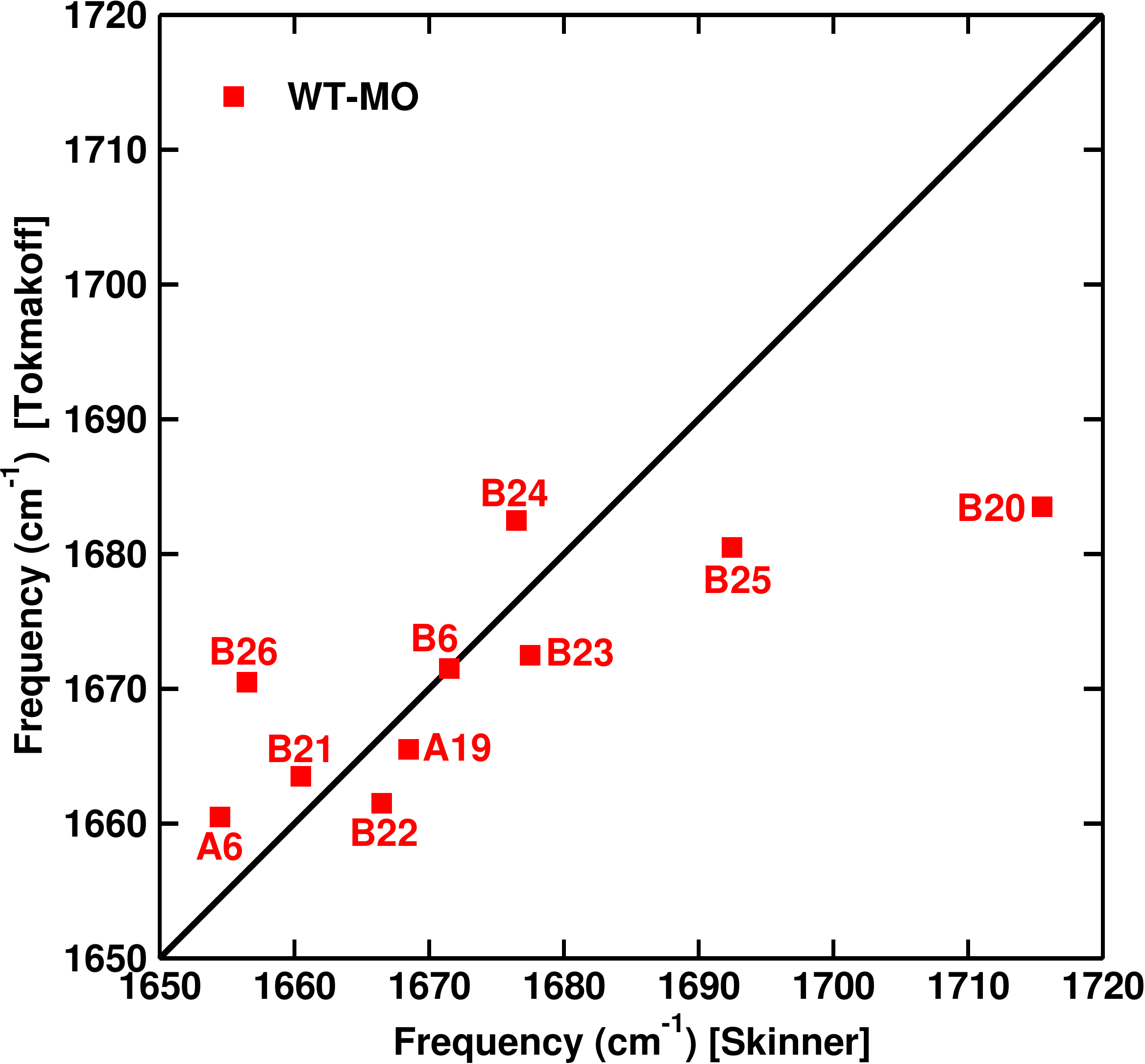}
\caption{Comparison between maximum frequency of 1D IR spectra for
  residues (A6, A19, B6, B20-B26, C6, C19, D6, D20-D26) based on two
  different maps\cite{skinner-map-JPCB2011,Tokmakoff-map-jcp-2013} for
  WT monomer. Snapshots from the same trajectory, run with constrained
  CO, were analyzed.}
\label{sifig:comp.map}
\end{center}
\end{figure}

\section{Validations for FFCFs}
\begin{figure}[H]
\begin{center}
\includegraphics[width=0.75\textwidth]{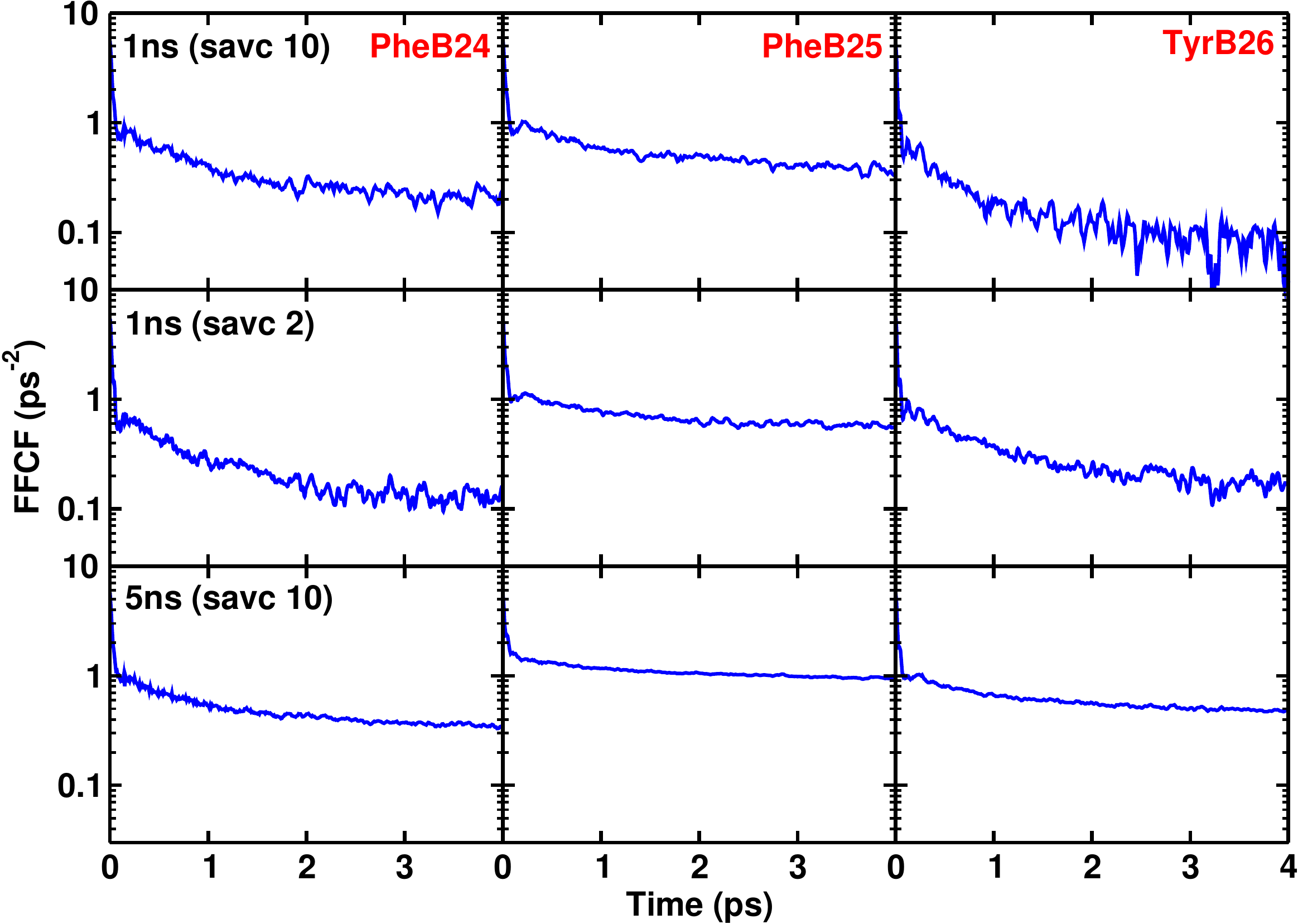}
\caption{FFCF for residues at the dimerization interface (B24-B26)
  from frequency trajectories based on "scan" for WT monomer. The FFCF
  is shown based on different simulation lengths (1 ns and 5 ns) and
  computing frequencies from snapshots saved every 2 or 10 fs (savc2
  and savc10). The overall shape of the FFCFs changes little whereas
  the noise level decreases especially for longer simulation
  times. Also, the magnitude of the static component increases for
  longer simulation times.}
\label{sifig:ffcf.test}
\end{center}
\end{figure}

\begin{figure}[H]
\begin{center}
\includegraphics[width=0.7\textwidth]{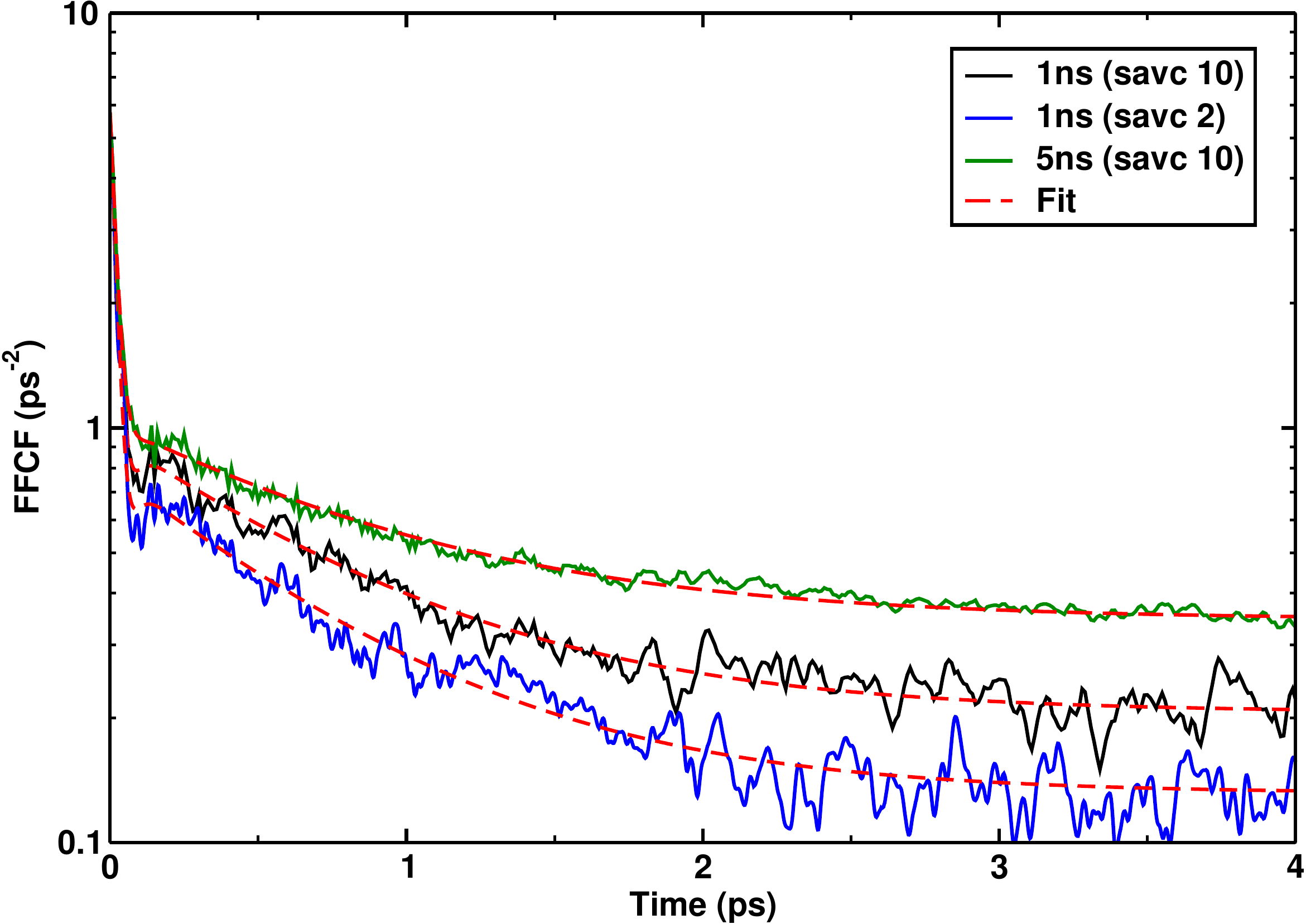}
\caption{Fitting the FFCF for residue B24 for the three analyses from
  Figure \eqref{sifig:ffcf.test}. The fitting parameters for the
  different FFCFs are summarized in Table \ref{sitab:ffcf.test}.}
\label{sifig:ffcf.b24}
\end{center}
\end{figure}

\begin{table}[h!]
\begin{tabular}{c|rrr|rr|r}
&$a_{1}$ & $\gamma$ &$\tau_{1}$ &$a_{2}$ &$\tau_{2}$ & $\Delta_{0}$\\
\hline
\textbf{1ns (savc10)}& & & & & &\\
B24 &4.64 &25.44 &0.025 & 0.75&0.74 &0.21\\
B25 &4.94 &22.19 &0.028 & 0.66&0.98  &0.37\\
B26 &4.97 &21.82 &0.019 & 0.62&0.62  &0.07\\
\hline
\textbf{1ns (savc2)} & & & & & &\\
B24 &4.64 &25.77 &0.023 & 0.65&0.68  &0.13\\
B25 &4.40 &20.96 &0.025 & 0.62&1.05  &0.54\\
B26 &4.89 &25.63 &0.020 & 0.77&0.70  &0.17\\
\hline
\textbf{5ns (savc10)} & & & & & &\\
B24 &4.74 &17.46 &0.023 & 0.69&0.83 &0.35\\
B25 &4.24 &0.00 &0.022 & 0.60&1.07  &0.94\\
B26 &4.96 &16.09 &0.021 & 0.59&0.92  &0.47\\
\hline
\end{tabular}
\caption{Parameters obtained from fitting the FFCF to
  Eq. 5 from ``scan'' frequencies for residues (B24-B26)
  for WT monomer based on different simulation length and different
  time separations between coordinates analyzed (every 2 or 10 fs -
  nsavc2 and nsavc10). The amplitudes $a_1$ to $a_3$ are in ps$^{-2}$,
  the decay times $\tau_1$ to $\tau_3$ in ps, the parameter $\gamma$
  in ps$^{-1}$, and the offset $\Delta_0$ in ps$^{-2}$.}
\label{sitab:ffcf.test}
\end{table}

\begin{table}[h!]
%\scriptsize{
%\centering\ra{1.3}
\begin{tabular}{c|rrr|rr|r|rr}
&$a_{1}$ & $\gamma$ &$\tau_{1}$ &$a_{2}$ &$\tau_{2}$ &$\Delta_{0}$ &$a_{3}$ &$\tau_{3}$\\
\hline
\textbf{WT monomer}\\
B24 &4.80 &27.19 &0.026 & 0.69&0.75 &0.21&&\\
B25 &4.66 &22.60 &0.030 & 0.62&1.02  &0.33\\
B26 &4.58 &22.32 &0.020 & 0.53&0.68  &0.06\\
\hline
\textbf{WT dimer M1} & & & & & & \\
B24 &4.65 &15.90 &0.070& 0.24 &3.80  &0.15  \\
B25 &3.62 &29.25 &0.026 &0.45 &1.19  &0.15\\
B26 &4.03 &17.16 &0.039 &0.40 &2.47  &0.50 \\
\hline
\textbf{WT dimer M2} & & & & & & \\
D24 &0.34 &17.51 &0.54 &3.62 &0.039  &0.14 &0.19&3.96\\
D25 &2.98 & &0.037 &0.41 &2.56  &1.67 \\
D26 &5.13 &14.17 &0.043 &0.42 &2.09  &0.27  \\
\hline
\hline
\textbf{B24A monomer} & & & & & & \\
B24 &4.81 &28.91 &0.021 &0.50 &0.60  &0.06\\
B25 &4.10 &15.70 &0.021 &0.64 &0.71  &0.11 \\
B26 &3.87 &25.00 &0.030 &0.60 &1.43  &0.52 \\
\hline
\textbf{B24A dimer M1} & & & & & & \\
B24 &4.81 &14.85 &0.050 &0.32 &1.46  & 0.43 \\
B25 &2.98 & &0.029 &0.63 &1.82  &0.70 \\
B26 &2.03 & &0.041 &0.27 &1.78  &0.44\\
\hline
\textbf{B24A dimer M2} & & & & & & \\
D24 & 1.19 & &0.043 &0.29 &1.24  &0.43 \\
D25 & 1.28 & &0.035 &0.29 &1.13  &0.33 \\
D26 & 4.67 &15.50 &0.043 &0.56 &1.46 &0.46 \\
\hline
\hline
\textbf{B24G monomer} & & & & & & \\
B24 &4.57 &29.61 &0.033 &0.41 &1.31 &0.27\\
B25 &4.34 &17.06 &0.019 &0.56 &0.80 &0.27  \\
B26 &4.41 &25.09 &0.022 &0.53 &0.51 &0.17&0.37&4.30\\
\hline
\textbf{B24G dimer M1} & & & & & & \\
B24 &1.28 & &0.032 &0.19 &1.21  &0.32   \\
B25 &3.55 & &0.019 &0.48 &1.19  &0.23 \\
B26 &3.71 & &0.030 &0.69 &2.05 &0.93\\
\hline
\textbf{B24G dimer M2} & & & & & & \\
D24 &2.46 &38.53 &0.015 &0.30 &1.22  &0.14\\
D25 &1.31& &0.034 &0.23 &1.82  &0.62  \\
D26 &3.94 &5.32 &0.042 &0.27 &2.14  &0.23 \\
\end{tabular}
\caption{Parameters obtained from fitting the FFCF to
  Eq. 5 for frequencies from INM for residues at the
  dimerization interface (B24-B26 and D24-D26). The amplitudes $a_1$
  to $a_3$ in ps$^{-2}$, the decay times $\tau_1$ to $\tau_3$ in ps,
  the parameter $\gamma$ in ps$^{-1}$, and the offset $\Delta_0$ in
  ps$^{-2}$.}
\label{sitab:ffcf.inm}
\end{table}

\bibliography{ins}